\documentclass[twocolumn, twocolappendix]{aastex631}
\shorttitle{GAMA: Stellar vs Dynamical Mass}
\usepackage{graphicx}	
\usepackage{amsmath}	
\usepackage{amssymb}	
\usepackage{xcolor, bm, booktabs}
\usepackage{multirow, blkarray}

\begin{document}

\title{Galaxy And Mass Assembly (GAMA): Stellar-to-Dynamical Mass Relation I. Constraining the Precision of Stellar Mass Estimates}

\correspondingauthor{M. Burak Dogruel}
\email{bdogruel@swin.edu.au}

\author[0000-0002-8688-4331]{M. Burak Dogruel}
\affiliation{Centre for Astrophysics and Supercomputing, Swinburne University of Technology, Hawthorn, VIC 3122, Australia}

\author[0000-0002-3958-0343]{Edward N. Taylor}
\affiliation{Centre for Astrophysics and Supercomputing, Swinburne University of Technology, Hawthorn, VIC 3122, Australia}

\author[0000-0002-9871-6490]{Michelle Cluver}
\affiliation{Centre for Astrophysics and Supercomputing, Swinburne University of Technology, Hawthorn, VIC 3122, Australia}

\author[0000-0003-2388-8172]{Francesco D’Eugenio}
\affiliation{Kavli Institute for Cosmology, University of Cambridge, Madingley
Road, Cambridge, CB3 0HA, United Kingdom}
\affiliation{Cavendish Laboratory - Astrophysics Group, University of Cambridge,
19 JJ Thomson Avenue, Cambridge, CB3 0HE, United Kingdom}

\author[0000-0002-2380-9801]{Anna de Graaff}
\affiliation{Max-Planck-Institut f\"ur Astronomie, K\"onigstuhl 17, D-69117, Heidelberg, Germany}

\author[0000-0001-9552-8075]{Matthew Colless}
\affiliation{Research School of Astronomy and Astrophysics, Australian National University, Canberra, ACT 2611, Australia}
\affiliation{ARC Centre of Excellence for All Sky Astrophysics in 3 Dimensions (ASTRO 3D), Canberra, ACT 261, Australia}

\author[0000-0002-6061-5977]{Alessandro Sonnenfeld}
\affiliation{Department of Astronomy, School of Physics and Astronomy, Shanghai Jiao Tong University, Shanghai 200240, China}


\begin{abstract}

In this empirical work, we aim to quantify the systematic uncertainties in stellar mass $(M_\star)$ estimates made from spectral energy distribution (SED) fitting through stellar population synthesis (SPS), for galaxies in the local Universe, by using the dynamical mass $(M_\text{dyn})$ estimator as an SED-independent check on stellar mass. We first construct a statistical model of the high dimensional space of galaxy properties; size $(R_e)$, velocity dispersion $(\sigma_e)$, surface brightness $(I_e)$, mass-to-light ratio $(M_\star/L)$, rest-frame colour, S\'ersic index $(n)$ and dynamical mass $(M_\text{dyn})$; accounting for selection effects and covariant errors. We disentangle the correlations among galaxy properties and find that the variation in $M_\star/M_\text{dyn}$ is driven by $\sigma_e$, S\'ersic index and colour. We use these parameters to calibrate an SED-independent $M_\star$ estimator, $\hat{M}_\star$. We find the random scatter of the relation $M_\star-\hat{M}_\star$ to be $0.108\text{dex}$ and $0.147\text{dex}$ for quiescent and star-forming galaxies respectively. Finally, we inspect the residuals as a function of SPS parameters (dust, age, metallicity, star formation rate) and spectral indices (H$\alpha$, H$\delta$, $D_n4000)$. For quiescent galaxies, $\sim65\%$ of the scatter can be explained by the uncertainty in SPS parameters, with dust and age being the largest sources of uncertainty. For star-forming galaxies, while age and metallicity are the leading factors, SPS parameters account for only $\sim13\%$ of the scatter. These results leave us with remaining unmodelled scatters of $0.055\text{dex}$ and $0.122\text{dex}$ for quiescent and star-forming galaxies respectively. This can be interpreted as a conservative limit on the precision in $M_\star$ that can be achieved via simple SPS-modelling.

\end{abstract}

\keywords{}


\section{Introduction}\label{sec:intro}

Stellar mass, $M_\star$, plays a key role in studies of galaxy formation and evolution, not least because so many galaxy properties are seen to correlate closely with mass.
However, stellar mass is not a direct observable and its estimations rely on mass-to-light ratio $(M_\star/L)$ determined by fitting the spectral energy distributions of the galaxies with stellar population synthesis (SPS) models \citep[][and references therein]{conroy2013}. 
These same SPS fits are also commonly used to derive parametric descriptions of other stellar population properties, including mean stellar age, metallicity, dust content, star formation rate (SFR), and more \citep[e.g., MAGPHYS;][FSPS; \citealt{conroy2009}, ProSpect; \citealt{robotham2020}, Prospector; \citealt{johnson2021}]{dacunha2008}.

Practically every aspect of SPS-modelling is subject to random and---more importantly---systematic uncertainties, which inevitably propagate through to the final results \citep[e.g.,][]{conroy2010a}. The choice of stellar initial mass function \citep[IMF, see review, e.g.,][]{hopkins2018} is usually taken as the single largest source of systematic uncertainty, though different choices of IMF can be approximately accounted for through a global scaling of the masses, similar to the Hubble parameter, $h$. Variations in IMF \citep[e.g.][and see review, \citealt{smith2020}]{baldry2003, vandokkum2008, guna2011} are possible, yet essentially impossible to implement within the usual SPS fitting frameworks. Further, the most prominent uncertainties in stellar evolution models include thermally pulsating asymptotic giant branch stars \citep[TP-AGBs; see, ][]{maraston2006, bruzual2007}, blue stragglers, the horizontal branch, and binary systems \citep[e.g.,][]{yi2003, lee2007, conroy2009}. 

The simplifying assumptions which make the SPS-modelling tractable also create potential sources of systematic uncertainties. All stars are typically assigned a single and uniform metallicity, regardless of their stellar age or star formation histories \citep[SFHs, but see][]{bellstedt2021}. This is further aggravated because of the coarse grid of metallicities covered in the spectral libraries \citep[see, e.g.][]{taylor2011}. The chosen priors on SFHs, and especially how SFHs are parameterised, can also have significant systematic effects on the derived stellar masses, as well as ancilliary parameters like SFRs, mean stellar ages, etc. \citep[e.g.,][]{carnall2019, leja2019, robotham2020}. 
Many, if not all, of these issues are long-standing and remain unresolved despite decades of efforts.

Considering the pivotal role of stellar masses in galaxy formation and evolution studies, and the potential of significant systematic errors or biases in these estimates, it is thus highly worthwhile to find some other SED-independent standard which can be used for quality assessment and/or calibration of the SPS-derived stellar mass estimates, including the random and systematic error budgets. What is required is a pure gravitational measure of stellar mass.

Strong gravitational lensing arguably provides the purest gravitational measure of stellar mass \citep[e.g.][]{posacki2015, smith2015, sonnenfeld2019}. However, these lensing events are rare, therefore such mass measurements are limited by small sample sizes as well as complex and potentially significant selection effects \citep[][]{sonnenfeld2023}.

A more common approach is to derive dynamical mass-to-light ratios $(M_\text{dyn}/L)$ for galaxy discs and/or bulges using the detailed dynamical modelling of integral field spectroscopy, IFS, \citep[e.g., Jeans Anisotropic Multi-Gaussian Expansion method -- JAM; ][]{cappellari2013}, for representative galaxy samples from IFS surveys like ATLAS3D, SAMI, MANGA, etc.

In this paper, we consider what can be done using single-fibre spectroscopy, as is the case in large scale galaxy redshift surveys such as SDSS, GAMA, DESI-BGS, and the 4MOST Hemisphere Survey (4HS). Relative to alternatives, the strength of single-fibre studies is far greater statistical power (with sample sizes in the millions for DESI-BGS and 4HS, cf.\ $\sim 15000$ for Hector) to disentangle primary correlations from the spurious ones, at the cost of detail and/or precision in the mass estimates for individual galaxies.

A simple dynamical mass estimator can be constructed via dimensional analysis as,
\begin{equation}
    M_\text{dyn}=k\frac{\langle V^2 \rangle R}{G},
    \label{eq:mdyn}
\end{equation}
where $\langle V^2 \rangle$ is ideally the mass-weighted and 3D-averaged velocity measured within some aperture and $R$ is typically taken to be the effective or half-light radius (within which half of the luminosity is emitted). Certainly, this combination of observable quantities is analogous to the virial theorem which holds for an ensemble of stars assumed to have regular orbits with randomly distributed phases (i.e., time averaged orbits). Some model or calibration is still needed to determine the so-called \textit{`virial coefficient'}, $k\sim 4$--6, also referred to as the \textit{`degree of virialization'}. For example, following the \cite{ciotti1997} approach, \cite{bertin2002} derive an approximate analytical expression for the virial coefficient as a function of S\'ersic index, $k(n)$, using simple dynamical models (with spherical, isotropic and non-rotating mass distribution) based on single stellar populations that follow a S\'ersic profile $(R^{1/n})$:
\begin{equation}
    k(n) \approx \frac{73.32}{10.465+(n-0.94)^2}+0.954 ~. \label{eq:kvn}
\end{equation}
On the other hand, \cite{cappellari2006} provide an empirical expression for $k(n)$ derived from detailed modelling of both imaging and spatially resolved spectroscopy for early-type galaxies from SAURON:
\begin{equation}
    k(n) = 8.87 - 0.831n + 0.0241n^2 ~ .
    \label{eq:bn}    
\end{equation}

With single-fibre spectroscopy, we typically use the 1D line-of-sight velocity dispersion $(\sigma)$ measured in the fibre aperture as the observable proxy for $\langle V^2\rangle$. 
Depending on the inclination or viewing angle, $\sigma$ encapsulates both the ordered rotation within the discs and the randomly oriented orbits in a bulge or halo. 
For early-type galaxies, \cite{cappellari2006, cappellari2013} empirically show that a simple dynamical mass estimate using the 1D velocity dispersion does indeed provide a reliable estimate of enclosed mass, insofar as it matches the more detailed JAM modelling results, $M_\mathrm{JAM}$.

Further, \cite{wel2022} use LEGA-C data and calibrate an empirical description of the relative contribution of ordered and disordered motion for inclined discs in terms of the observed axis ratio $(q=b/a)$ as,
\begin{equation}
    V^2 = k(q) \sigma^2 : k(q) \approx \left(0.87 + 0.38 e^{-3.78(1-q)}\right)^2~ ,
    \label{eq:kq}
\end{equation}
which can be implemented in equation~(\ref{eq:mdyn}) by replacing $k$ with $k(n)k(q)$.

The purpose of this paper is to quantify the empirical relation between simple $M_\star$ and $M_\text{dyn}$ estimates, and in particular to separate out the statistical correlations with a number of other observable galaxy properties which trace different aspects of galaxy phenomenology. The hope is that quantifying these correlations might enable us to identify whether or where there are clear and large discrepancies between $M_\star$ and $M_\text{dyn}$ which might point to systematic errors in one or the other estimator. 

More specifically, our goals are:
\begin{enumerate}
    \item to quantify the direct correlation between simple $M_\star$ and $M_\mathrm{dyn}$ estimates for galaxies, including both the mean relation and the intrinsic scatter of the relation; \label{item:goal1}
    \item to quantify the correlations between $M_\star/M_\mathrm{dyn}$ and a number of observable galaxy properties, including what variance and correlation can be uniquely attributed to each of these properties; \label{item:goal2}
    \item to measure the plausible error budget for both $M_\star$ and $M_\mathrm{dyn}$, including random and systematic errors that can potentially be linked to, e.g., IMF variations. \label{item:goal3}
\end{enumerate}

This paper is structured as follows:
We describe the GAMA data that we have used, including sample selection and quality control, in section \ref{sec:data}. We provide a description of our method in section \ref{sec:method}. Our results and their implications are presented in three sections: we first present the results regarding the relation between $M_\star$ and $M_\mathrm{dyn}$ deduced from this framework in section \ref{sec:results}, we then show which galaxy properties are the main drivers of the variation in $M_\star/M_\mathrm{dyn}$ ratio in section \ref{sec:trends_mstar_to_mdyn} and lastly we present the third set of our results, pertaining to testing the precision of SED-derived $M_\star$ in section \ref{sec:mstar_precision}. Finally, we summarise our findings in section \ref{sec:summary}.

We adopt a concordance cosmology with $H_0=100h$ km/s/Mpc, $\Omega_\Lambda=0.7$ and $\Omega_m=0.3$ for conversion from angular radius $(\theta'')$ to physical radius $(R_e/\text{kpc}h^{-1})$, whereas, the stellar mass estimates that we use assume a \cite{chabrier2003} IMF and a concordance cosmology with $h=0.7$.

\section{Data}\label{sec:data}

This work is based on a mass limited sample of galaxies drawn from the GAMA survey \citep[][]{driver2011, liske2015, driver2022}. GAMA combines a spectroscopic redshift survey undertaken using the 3.9m Anglo-Australian Telescope (AAT) with multiwavelength survey across 5 sky regions with a 250deg$^2$ coverage of equatorial and southern sky. The latest and final data release is described by \cite{driver2022} and it includes $\sim$300 000 spectroscopic redshift measurements achieving a remarkable completeness level of 95 per cent for a magnitude limit $m_r\leqslant19.65$ mag \citep[][]{bellstedt2020a}.

The principal value of GAMA for this study is that it provides all of the necessary quantities; namely: multiband SEDs; SPS-derived stellar mass estimates and ancillary stellar population parameters; S\'ersic-fit structural parameters; and central velocity dispersions. We discuss each of these data products in turn below, before describing our specific sample definition in section \ref{sec:data_sample}.

\subsection{Optical-to-NIR SEDs and stellar masses}\label{sec:data_phot_mstar}

As outlined in \cite{bellstedt2020a, bellstedt2020b}, GAMA provides photometry in 19 bands from far-UV to the IR, performed with {\large{P}}RO{\large{F}}OUND package \citep[][]{robotham2018}. In brief, source detection and deblending is done based on combined images from KiDS $r-$band and VIKING Z-band, i.e., $r+$Z. Definitions of initial isophotal outlines (ProFound segments) from the $r+$Z detection image are then iteratively dilated to obtain a kind of curve-of-growth total flux measurement from the segments, while ensuring that neighbouring segments/apertures do not overlap. The resultant FUV-IR Spectral Energy Distributions (SEDs) are corrected for foreground Galactic extinction using the Planck $E(B-V)$ maps \citet{planck2013} and matched back to spectroscopy and redshifts \citep[DMU: \textit{gkvInputCatv02},][]{bellstedt2020a}, which gives the basis for SPS-modelling for stellar masses and ancillary stellar population parameters. To ensure consistency between the SPS-derived stellar mass estimates and the S\'ersic fits, we re-normalise the SEDs to match the S\'ersic $Z$-band magnitude. In other words, we use the multiband ProFound photometry for colours (and hence stellar populations, $M_\star/L$, etc.) and use the S\'ersic photometry for total magnitude (and hence $M_\star$).

The stellar mass estimates that we use are derived following \cite{taylor2011}: namely, using SPS models from \cite{bruzual2003}; the IMF of \cite{chabrier2003}; exponentially declining SFHs; and single screen dust with the \cite{calzetti2000} attenuation law. In the fits, the different bands are weighted to consider an approximately fixed rest-frame wavelength range of $3000-11000$ \AA. Stellar mass and other parameter estimates are done in a Bayesian way (i.e., marginalising over a fixed grid in parameter space, with plausible priors), which encapsulates the important covariances and degeneracies in the modelling. 

Finally, we add the more sophisticated stellar mass estimates derived using the SED-fitting code {\sc ProSpect} \citep[][]{robotham2020} which allows for gas-phase metallicity $(Z_\text{gas})$ varying with time and in turn makes possible to process more complicated SFHs. Notably, as examined in \cite{robotham2020} and mentioned in \cite{driver2022}, {\sc ProSpect} masses are systematically 0.06 dex larger than the ones derived by \cite{taylor2011} used in this study.

\subsection{S\'ersic Modelling}\label{sec:sersic}

To characterise the total flux, effective radii, and S\'ersic index for the galaxies in our sample, we use the S\'ersic-fit values derived for the VIKING-Z band, which provides a deeper probe in near infrared with a magnitude limit of $\sim 23$ mag (AB) and $1\arcsec$ of seeing \citep[see][Table 1]{bellstedt2020a}. The process and pipeline for the single-S\'ersic fits that we use is presented by \cite{kelvin2012}. Galaxy sizes, i.e. effective radii $(\theta_e'')$, and S\'ersic indices $(n)$ are based on 2D S\'ersic model fitting to the surface brightness profile, with models truncated at 10 times the angular effective radius to prevent extrapolating flux into the uncertain regime of large radii. While these magnitudes are in agreement with Kron or Petrosian aperture based photometry up to $n\sim 3$, they recover an additional $\sim 0.2$ mag in high S\'ersic index galaxies \citep[][]{kelvin2012}, as expected. 

One important issue is that the parameters in the S\'ersic fits can be strongly covariant: S\'ersic index, total magnitude estimate and effective radius are interdependent as discussed in detail by \cite{kelvin2012}. While the catalogues do not include the correlation coefficients that describe this covariance, it is possible to determine these values empirically using independent measurements in neighbouring bands, as described in Appendix \ref{sec:calc_coverrors}.

\subsection{Velocity dispersions and dynamical masses}\label{sec:veldisp_mdyn}

The GAMA spectroscopic database incorporates previously obtained spectra and redshifts from multiple sources, including SDSS \citep[][]{ahn2014}, 2dFGRS \citep[][]{colless2001_2df} and 6dFGS \citep[][]{jones2004, jones2009}, with some targets in these surveys having been observed again by the GAMA team. This heterogeneous spectroscopic data actually presents a challenge in obtaining coherent and consistent velocity dispersion measurements for targets that have spectroscopic data obtained from different instruments and with different data reduction pipelines.

For GAMA DR4, we have used {\sc pPXF} \citep[][]{cappellari2017} with templates from the stellar spectra library MILES \citep[][]{sanchez2006} having a resolution of 2.51 \AA \citep[][]{falcon2011} to derive velocity dispersion measurements for all spectra in the GAMA database. These measurements, which are given in the DMU \textit{VelocityDispersionsv01}, are derived by fitting the 1D spectra as a non-negative linear superposition of a set of stellar template spectra, and assuming a Gaussian line-of-sight velocity distribution.

To ensure the coherence and consistency of our measurements, we re-calibrate the results from each independent survey through cross-survey comparison. Specifically, we re-calibrate the velocity dispersions from all other surveys to match those from SDSS, partly because SDSS has the best spectral resolution \citep[see][Table 2]{driver2022} and also provides a connection to SDSS peculiar velocity measurements \cite[][]{howlett2022} which forms the basis of our next companion paper. Furthermore, we validate/calibrate the random errors in the measurements through comparisons of repeat observations of the same target within a single survey (i.e., intra-survey comparison). We find that this approach works well for the 6dFGS and GAMA data, but not for 2dFGRS, where the process results in larger random and systematic errors. Accordingly, we only use velocity dispersions measured from the 6dFGS, SDSS and GAMA spectroscopy. We defer further details of cross-survey and intra-survey comparisons and calibration to a future paper.

Finally, we combine these velocity dispersion measurements with the VIKING-Z band effective radii derived through S\'ersic fits (section \ref{sec:sersic}), and use the \cite{bertin2002} prescription for $k(n)$ to calculate our simple dynamical mass estimates. This choice is mainly because the empirical relation for $k(n)$ of \cite{cappellari2006} is derived based on only early-type galaxies, while we consider both early and late-type galaxies in this study. However, we also use the $M_\mathrm{dyn}$ derived using equation (\ref{eq:bn}) to study the relation between $M_\star$ and $M_\mathrm{dyn}$, then compare the results to the ones from when equation (\ref{eq:kvn}) is used.

\subsection{Data limits and sample selection}\label{sec:data_sample}

Our basic sample selection is $\log M_\star/M_\odot > 10.3$ and $0.01<z<0.12$. We adopt the following simple quality control cuts for our sample to ensure our $M_\star$ and $M_\mathrm{dyn}$ estimates are robust and comparable.

First, to ensure reliable spectroscopic redshifts, we require that targets have a GAMA redshift quality flag $nQ\geqslant 3$, corresponding to a redshift confidence $>90\%$ \citep[see, equation 15 of][]{liske2015}. Next, we discard cases where the S\'ersic fits have failed or have given extreme values, by requiring that the S\'ersic indices are within the range $0.3 < n < 10$ to be included in the sample. Additionally, we discard cases where the total flux derived from the S\'ersic fits differs from the iteratively dilated {\sc ProFound} segment fluxe by more than 0.5 dex, which would imply an overall inconsistency between the SED-aperture and the S\'ersic fits. As described in \ref{sec:veldisp_mdyn}, we only include velocity dispersions measured from 6dFGS, SDSS and GAMA. Finally, we require the velocity dispersions in the range $60 < \sigma_e~\mathrm{[km/s]} < 450$ with estimated measurement uncertainties $\varepsilon_\sigma < 0.25\sigma_e+25$. 

\begin{figure*}
    \includegraphics[width=\textwidth]{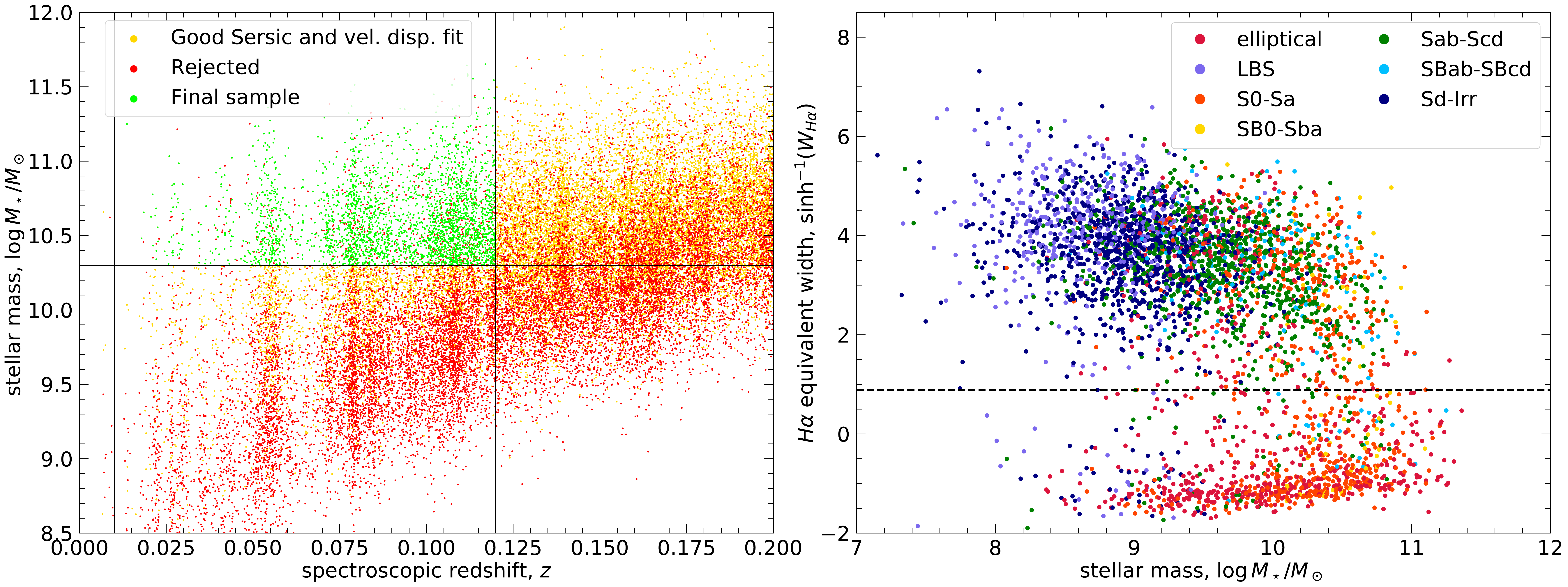}
    \caption{\textit{Left}: GAMA sample selection. The stellar mass and redshift distribution of the galaxies from GAMA obtained by using only reliable redshifts $(nQ\geqslant 3)$. The orange coloured points represent the sample with good S\'ersic and velocity dispersion fits, whereas the red coloured points represent the sample rejected due to failed quality controls in these fits. The green points show the final sample with black lines showing the adopted redshift and stellar mass limits. \textit{Right:} Distinction between quiescent and star-forming galaxies. The distribution of stellar mass and equivalent width of the H$\alpha$ line, colour coded with respect to Hubble type, taken from the DMU \textit{VisualMorphologyv03} \citep[][]{kelvin2014}. Positive equivalent width values imply emission lines. The horizontal dashed line shows our selection criteria between quiescent (H$\alpha\text{EW}<1$\AA) and star-forming galaxies (H$\alpha\text{EW}\geqslant 1$\AA).}
    \label{fig:sample_selection}
\end{figure*}

The left panel of Figure \ref{fig:sample_selection} shows the distribution of stellar masses and redshifts, where the entire sample with reliable redshifts contains 74594 galaxies. Enforcing the quality controls for the S\'ersic and velocity dispersion fits leaves us a parent sample of 32707 galaxies, shown in orange. Finally, applying the stellar mass and redshift selection, the resulting sample contains 2850 galaxies in VIKING-ZYJHK bands and it is shown with green. 

Following the arguments of \cite{taylor2011, lange2015, baldry2018}, our redshift and mass selection result in a hypothetically volume limited sample. 

\subsection{Early/Late type galaxy selections}\label{sec:data_etgltg}

Due to the significant overlap of red and blue galaxy populations seen in e.g. colour-mass, colour-S\'ersic index diagrams \citep[e.g.][and many others]{blanton2009, taylor2015}, making an ultimate distinction between early and late type galaxies is not a trivial task. A S\'ersic cut with $n=2.5$ \citep[][]{shen2003} can be used where smaller $n$ values correspond to late-type galaxies (LTGs) and greater ones correspond to early-type galaxies (ETGs), as well as a colour cut, such as $g-i>0.65$ for ETGs, can be used \citep[][]{lange2015}. 

In this work, we make use of one of the main differences between the two populations, that is, star formation. While LTGs are star-forming (SF) galaxies, it has diminished for ETGs, thus making them quiescent (Q) galaxies. Star formation can be measured by, e.g., H$\alpha$ emission lines. Thus, H$\alpha$ line will be an absorption feature in the spectra of quiescent galaxies which means that the equivalent width (EW) will have negative values. As seen in the right panel of Figure \ref{fig:sample_selection}, H$\alpha$ EW provides a quite clear bimodality. Following Figure 1 of \cite{taylor2015}, we separate our sample with H$\alpha$ EW $\geqslant1$\AA$\,$ as SF and H$\alpha$ EW $<1$\AA$\,$as Q galaxies. Here, we take H$\alpha$ and H$\delta$ equivalent widths from the DMU \textit{SpecLineSFRv05} \citep[][]{gordon2017}.

We present our final sample in Figure \ref{fig:fitcorner}, showing the distributions of galaxy properties $r\equiv\log R_e~[h^{-1}\text{kpc}]$, $s\equiv\log\sigma_e~\mathrm{[km/s]}$, $i\equiv \log\langle I_e\rangle[L_\odot/\text{pc}^2]$, $m^*\equiv\log M_\star[M_\odot]$, $\ell\equiv\log L[L_\odot]$, $c=(g-i)_\text{rest}$ rest-frame colour, $\nu\equiv\log\,n$ and $m^d\equiv\log M_\text{dyn}[M_\odot]$ among Q and SF galaxies.

\section{Method: Gaussian Based Hierarchical Bayesian Model}\label{sec:method}

\subsection{Motivation and statement of the problem}\label{sec:motivation}

While $M_\mathrm{dyn}$ measurements obtained through simple estimators are usually interpreted as referring to the total mass of the disc and bulge components of a galaxy (cf. the larger dark matter halo), there is a subtlety depending on how $k$ is calibrated. For example, \cite{ciotti1997} prescription oriented $k(n)$ of \cite{bertin2002} (equation~\ref{eq:kvn}) strictly applies to a purely stellar bulge, whereas, the \cite{cappellari2006} calibration (equation~\ref{eq:bn}) is derived from projecting S\'ersic profiles of 25 E/S0 galaxies to $1R_e$. In the latter case, any non-stellar mass component (e.g., molecular or atomic gas) will contribute directly to $M_\mathrm{dyn}$ if these components follow a similar density profile. However, if the density profile is different (e.g., the dark matter halo), the contribution from non-stellar mass changes depending on the amount and the distribution of that mass.

To put this in another way, the dynamical mass estimator in equation~(\ref{eq:mdyn}) is model dependent, thus it is appropriate to include a correction factor, $f_\mathrm{dyn}$, to account for possible random or systematic errors stemming from the shortcomings of the model. Defining a similar factor for $M_\star$ estimates as $f_\star$, we have
\begin{equation}
    \frac{M_\star}{M_\mathrm{dyn}} = \frac{f_\star M_\star}{f_\mathrm{dyn}M_\mathrm{dyn}} = \left( f_\star \frac{M_\star}{L} \right) \left( f_\mathrm{dyn} k\right)^{-1} \frac{GL}{\sigma_e^2 R_e} ~ .
    \label{eq:f_mstar_f_mdyn}
\end{equation}
Here, the correction factors $f_\star$ and $f_\mathrm{dyn}$ contain any number of contributing factors such as corrections to the IMF, dark matter fraction, the contribution of atomic/molecular gas and the modelling uncertainties in $M_\star/L$ and $k$. In general, unambiguously disentangling all possible contributions which are largely if not completely degenerate, is an arduous task.

Characterising the correspondence between simple estimates of stellar and dynamical masses will also lead us to explore potential systematic errors in $M_\star$ estimates by studying the residual trends in the stellar-to-dynamical mass ratio $(M_\star/M_\mathrm{dyn})$ as a function of stellar population parameters, a similar approach adopted by e.g., \cite{taylor2010, esdaile2021} and \cite{degraaff2021}.

For example, if there is some overall deficiency in the SPS-modelling, we might expect to see it in the form of systematic variations in $M_\star/M_\text{dyn}$ as a function of some stellar population parameter like (rest-frame) colour. In such a scenario, this systematic variation with colour could be interpreted as a measure of the correction factor, $f_\star$, and might even be considered in terms of a varying IMF. Similarly, if we see a systematic trend with S\'ersic index $(n)$, which is a proxy for structural differences among galaxies, it might be a sign of structure dependent deficiency in the modelling. Furthermore, we might expect to see the stellar-to-halo mass relation $(M_\star/M_\mathrm{halo})$ manifesting itself as a trend in $M_\star/M_\mathrm{dyn}$, which might further be used to derive a refined prescription for $k(n)$. These being said, we have to point out that we will not be able to break the fundamental degeneracy: an observed correlation between $M_\star/M_\mathrm{dyn}$ with, e.g., colour, S\'ersic index or $M_\star$ might imply differential errors in $M_\star$ through a varying IMF, or real variation in $M_\star/M_\mathrm{dyn}$ due to variations in gas content and/or dispersion in the $M_\star/M_\mathrm{halo}$ relation (i.e. the dark matter fraction), or any combination thereof.

\subsection{Gaussian forward modelling}\label{sec:gaussian_forward}

Considering that there are strong correlations among many galaxy properties, the principal concern here is uniquely and robustly identifying trends with particular observable properties. Many previous studies have adopted an approach in which they studied the residuals from a simple 2D fit to the $M_\star$--$M_\mathrm{dyn}$ relation (i.e., fitting $M_\star$ as a function of $M_\mathrm{dyn}$ or vice versa) plotted against other properties, hoping to identify any leading trends or sources of error.

In this work, we instead set out to obtain a complete statistical description of the full high dimensional distribution of galaxy properties, so that it makes possible to simultaneously and explicitly account for the elaborate web of real correlations among these properties. This approach enables us to uniquely quantify how much of the random and systematic errors in $M_\star$ and/or $M_\mathrm{dyn}$ can be directly attributed to different properties.

More specifically, we use a multivariate Gaussian model to describe the observed galaxy distribution in $M$-dimensional parameter space\footnote{This is in some sense similar to the approach of \cite{oldham2017} who simultaneously model multiple scaling relations among galaxy properties for a high redshift cluster.}. \cite{saglia2001} and \cite{colless2001} have first used a 3D Gaussian model to characterise the Fundamental Plane (FP) of early-type galaxies, showing how the non-negligible (covariant) measurement errors, explicit selection cuts, and large spread of selection weights (implicit selection cuts, e.g., magnitude limits) can be elegantly accounted for within a Gaussian framework. \cite{magoulas2012} have also shown in their comprehensive work that a 3D Gaussian maximum likelihood (ML) fitting is a statistically rigorous, unbiased method providing a remarkably successful description of the FP and demonstrated how this scheme is superior to simple regression methods. This approach has become the state of the art for FP studies, which have otherwise been troubled by systematic uncertainties stemming from the choice of fitting formalism/procedure. In this work, we generalise this 3D Gaussian ML method to higher dimensions so that it can simultaneously fit more galaxy parameters, and adapt it to a Bayesian framework.

The choice of Gaussian to represent the distribution of galaxies in the observable parameter space is wrong in detail, though that is not necessarily the point: the model does not have to be right to be useful. Even so, we still see (and show in Figure \ref{fig:fitcorner}) that a Gaussian model does indeed provide a reasonable description of the data, especially of the bivariate linear correlations that are the focus of our work here.

The motivation for adopting a Gaussian model is completely pragmatic. For the purposes of our study, the most important advantage that comes forward is that \textit{once we have the Gaussian description of the full parameter space, it is straightforward to derive any linear relation between the galaxy properties of interest}. This advantage stems partly from the facts that 1) any slice through a multivariate Gaussian distribution is itself Gaussian, and 2) any linear combination of independent Gaussian random variables is itself Gaussian. The other aspect of this is that commonly used linear regression methods like Ordinary Least Squares (OLS) and Orthogonal Distance Regression (ODR) can be expressed in terms of the Gaussian parameters that define the covariance matrix. In other words, our Gaussian model can be seen as just a tool to design a powerful and convenient way to package all possible linear fits within the $M$-dimensional parameter space that we consider.

In Appendix \ref{sec:analytical_gaussian}, we lay out the basic mathematical foundations of our Gaussian model in a Bayesian framework, which includes the details on: handling the error propagation while account for the possible covariances among them (Appendix \ref{sec:cov_errs}), accounting for the selection effects (Appendix \ref{sec:selection_effects}), and implementing a scheme for the identification and rejection of outliers (Appendix \ref{sec:mixture}). Then, we describe how we make the modelling problem tractable through using the No U-Turn Sampler \citep[\textit{NUTS},][]{nuts} algorithm implemented in \textit{STAN} \citep[][]{standevteam}, which makes Markov Chain Monte Carlo (MCMC) sampling some orders of magnitudes faster (Appendix \ref{sec:stan}), and finally, we describe the relation between our Bayesian approach to Gaussian forward modelling and the standard frequentist approach to regression, namely, OLS and ODR (Appendix \ref{sec:lintransform}). 

\section{Results I: Relation Between Stellar and Dynamical Masses}\label{sec:results}

In this section, we study the consistency of the relation between the two different mass estimates; $M_\star$ and $M_\text{dyn}$, and quantify its strength and tightness using the metrics correlation coefficient $(\rho)$ and intrinsic scatter $(\sigma_\text{int})$, respectively. Our main goal here is to find the combination of parameters that give us the strongest (high $\rho$) and tightest (low $\sigma_\text{int}$) relation. This will allow us to examine which parameters are the most likely ones to drive the difference between these mass estimates and their implications for different galaxy populations.

We achieve this task by making use of the other advantageous feature of Gaussian approach, which allows the linear correlations to be easily derived as outlined in Appendix \ref{sec:lintransform}. The rest of this section is structured as follows: in section \ref{sec:ndgalaxies}, we give the \textit{global} result of our model of the 8D parameter space for both galaxy types, which can be perceived as the \textit{parent} model and forms the basis of our analysis for the rest of the paper. In section \ref{sec:mstar_vs_mdyn}, we first show which method of linear regression is suitable for different purposes (section \ref{sec:linreg}), then we show the intrinsic relation between $M_\star$ and $M_\text{dyn}$ derived from conditional distributions which make it possible to isolate other observable galaxy properties (section \ref{sec:marg_vs_cond}).

\subsection{The M-dimensional parameter space of galaxies}\label{sec:ndgalaxies}

\begin{figure*}
    \centering
    \includegraphics[width=\textwidth]{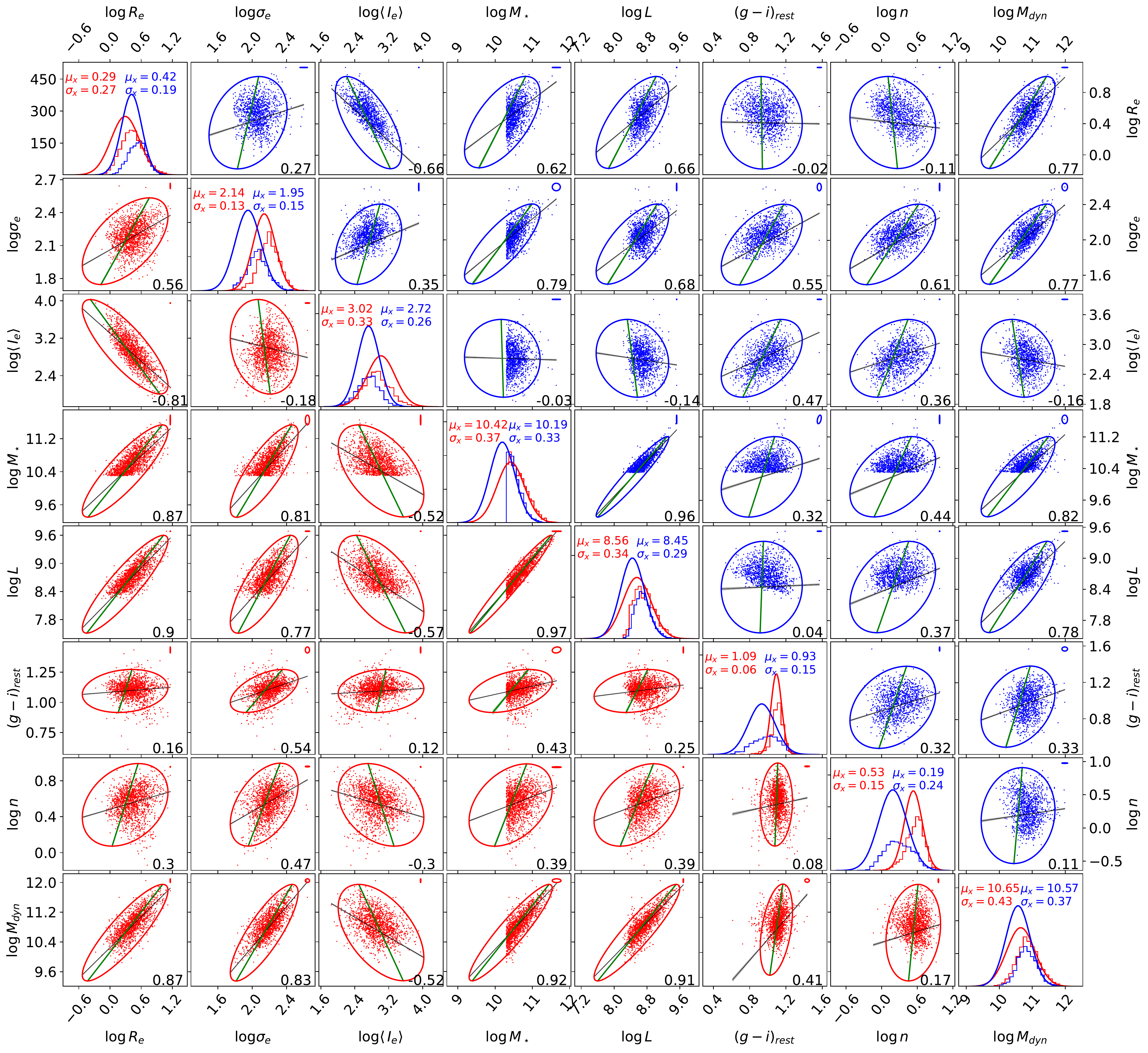}
    \caption{The 8D parameter space of galaxies in the GAMA sample, from left to right: effective radius, velocity dispersion within the effective radius, effective surface brightness, stellar mass, luminosity, rest-frame colour, S\'ersic index and dynamical mass. Figure shows both the data and a summary of the modelling results with red and blue colours representing Q and SF galaxies respectively. Diagonal panels show the 1D marginal distributions of each parameter in the data set. Smooth curves are the 1D projections of the 8D fit, accounting for selection effects. Mean $(\mu_x)$ and standard deviation $(\sigma_x)$ obtained from the fit are also given on the top of each diagonal panel. Off diagonal panels show the 2D projections with $3\sigma$ Gaussian ellipses from the fit. The lines show the forward, $y=f(x)$, and inverse, $x=f^{-1}(y)$, OLS fits derived from the 2D projections of the covariance matrix as described in section \ref{sec:lintransform} (slope $a=\rho\sigma_y/\sigma_x$ and intercept $b=\mu_y - a\mu_x$), while uncertainties in the fitted lines are too small to be visible in this scale. Here, the black lines show the forward fit, while green lines show the inverse fit. The median error ellipses are shown on the top right corner of each off-diagonal panel. Finally, pairwise correlation coefficients $(\rho)$ are given in the bottom right corners.}
    \label{fig:fitcorner}
\end{figure*}

Before moving on, we first present both the data and modelling results in Figure \ref{fig:fitcorner}. Diagonal panels show the 1D marginal distributions of each parameter in the data set: histograms show the data as observed, and the smooth curves show the Gaussian model description. The off-diagonal panels show the 2D projections of the 8D parameter space, where the upper and lower triangular parts are for SFs and Qs respectively. Within each of these panels, the points show the data as observed, with median error ellipses given in the top right corner.
The large coloured ellipses trace the 3$\sigma$ points of the underlying Gaussian model for each 2D projection of the 8D parameter space. Then, lines show the forward and inverse linear correlations between
each pair of parameters, as derived from the Gaussian fit parameters (see section \ref{sec:lintransform}).

This figure can also be read as a graphical table of our fitting results. In the diagonals, we give the means $(\mu)$ and standard deviations $(\sigma)$. In the off-diagonals, we give the correlation coefficients $(\rho)$. These values give a complete description of the 8D parameter space of galaxies, from which we can derive any other linear correlation. 

\subsection{The relation between stellar mass and dynamical mass}\label{sec:mstar_vs_mdyn}

\subsubsection{Linear regression and linear combinations}\label{sec:linreg}

As previously discussed many times in the literature \citep[e.g.][]{bernardi2003c, magoulas2012} the linear regression coefficients can significantly change with respect to the adopted regression method. OLS regression minimises the residuals in one of the variables, which makes it useful for predicting a parameter from another one (or from a combination of other parameters). The forward fit is the linear relation that best predicts $y$ as a function of $x$: $M_\star \propto M_\text{dyn}^\alpha$; and the inverse fit is $x$ as a function of $y$: $M_\text{dyn}\propto M_\star^\beta$. 

On the other hand, ODR provides the description of the underlying relation by minimising the residuals in the orthogonal direction, which is astrophysically the most interesting. This way, we can have insights on the structural differences between the galaxy populations that may cause the difference in these mass estimates.

The relation between $M_\star$ and $M_\text{dyn}$ can be decomposed in terms of observable galaxy properties as
\begin{equation}
    \frac{M_\star}{M_\text{dyn}} = \frac{2\pi G}{k(n)} \left( \frac{M_\star}{L_i} \right) \langle I_e \rangle R_e \sigma_e^{-2}.
    \label{eq:mstar_to_mdyn}
\end{equation}
Since $\log\,M_\star/L_i\propto (g-i)_\text{rest}$ and, as a first order approximation, 
\begin{equation}
    \log k(n) \propto 
    \begin{cases}
        \text{constant}, & n \lesssim 1.4 \\
        - \log n, & \text{otherwise}
    \end{cases}
\end{equation}
we can study the relation as, e.g., $(\bm{m^*}, \bm{m^d}\,|\, \bm{r}, \bm{s}, \bm{i}, \bm{\nu}, \bm{c})$. However, considering that our $M_\text{dyn}$ estimates are obtained through equation~(\ref{eq:mdyn}), the inclusion of $\bm{r}$, $\bm{s}$ and $\bm{i}$ at the same time will result in over-fitting of this relation. Moreover, under the assumption of homology, we get $\bm{m^d}=\bm{r+2s}+\text{constant}$ which causes linear dependence leading to a singular matrix. Therefore, we study a 5D relation by taking a subset of the 8D parameter space as $\bm{Y}=(\bm{m^*}, \bm{m^d}, \bm{s}, \bm{\nu}, \bm{c})$. This 5D relation can now be evaluated from the corresponding model $\bm{Y} \sim \mathcal{N}(\bm{\mu_Y, \bm{\Sigma_Y}})$ where the covariance matrix $\bm{\Sigma_Y}=\bm{J_Y \Sigma J_Y}^\intercal$ and the mean vector $\bm{\mu_Y}=\bm{J_Y\mu}$ are obtained from the parent model via
\begin{equation}
    \bm{J_Y} = \begin{blockarray}{*{9}{c}}
        & r & s & i & m^* & \ell & c & \nu & m^d \\
      \begin{block}{c(*{8}{c})}
        m^* & & & & 1 & & & & \\
        m^d & & & &  & & & & 1 \\
        s & & 1 & &  & & & & \\
        \nu & & & &  & & & 1 & \\
        c & & & & & & 1 & &  \\
      \end{block}
    \end{blockarray}    
\end{equation}
where the omitted elements are zero.

\subsubsection{Marginal versus conditional relations}\label{sec:marg_vs_cond}

The subspace $\bm{Y}$ can be partitioned as $\bm{Y} = (\bm{Y_a}, \bm{Y_b})$. Similarly, the corresponding mean vector and covariance matrix can also be partitioned in block form as
\begin{equation}
    \bm{\mu_Y} = (\bm{\mu_a}, \bm{\mu_b})\quad \text{and}\quad 
    \bm{\Sigma_Y}=\begin{pmatrix}
                    \bm{\Sigma_{aa}} & \bm{\Sigma_{ab}} \\
                    \bm{\Sigma_{ba}} & \bm{\Sigma_{bb}}
                \end{pmatrix}
    \label{eq:partition}
\end{equation}
where, $\bm{\Sigma_{ba}} = \bm{\Sigma_{ab}}^\intercal$. The marginal distribution of $\bm{Y_a}$ is $\bm{Y_a}\sim \mathcal{N}(\bm{\mu_a}, \bm{\Sigma_{aa}})$ and the conditional distribution of $\bm{Y_a}$ given $\bm{Y_b}=\bm{y_0}$ becomes $\bm{Y_a} | \bm{Y_b} \sim \mathcal{N}(\bm{\Tilde{\mu}}, \bm{\Tilde{\Sigma}})$ where,
\begin{align}
    \bm{\Tilde{\mu}} &= \bm{\mu_a} + \bm{\Sigma_{ab}\Sigma_{bb}^{-1}}\left(\bm{y_0}-\bm{\mu_b}\right)\nonumber\\
    \bm{\Tilde{\Sigma}} &= \bm{\Sigma_{aa}} - \bm{\Sigma_{ab}\Sigma_{bb}^{-1}\Sigma_{ab}}^\intercal.
    \label{eq:tilda}
\end{align}

It is crucial to emphasise here that fits from conditional distributions are more suitable to examine the dependence of $M_\star-M_\text{dyn}$ relation on other parameters, because, as per the name, everything else is marginalised out in marginal distributions. In that case, it is just the projection of the 8D parent model on the 2D $M_\star-M_\text{dyn}$ relation (in other words, fitting the $M_\star-M_\text{dyn}$ pair as a function of one another), in which strong correlations with other observable galaxy properties (e.g., $M_\star-R_e$, $M_\star-\sigma_e$, $R_e - \langle I_e \rangle$) are indistinguishably included. Using conditional distributions provides an elegant way to perform partial linear regression \citep[e.g.][and references therein]{oh2022, barsanti2022} which makes it possible to find the true correlation (or, as we call it, the intrinsic relation) between two properties, while isolating the cross-correlations from other properties.

We now examine the $M_\star-M_\text{dyn}$ relation separately for each galaxy population by inferring the coefficients, slope and intercept, from the conditional and marginal distributions, taking $\bm{Y_a} = (\bm{m^*}, \bm{m^d})$ and $\bm{Y_b}=(\bm{s}, \bm{\nu}, \bm{c})$. Here, using the relevant covariance matrices, we calculate both the coefficients that minimise the sum of the orthogonal distances (ODR) and the coefficients that minimise the residuals in the $y-$direction (parallel or direct fit: OLS). We denote the ODR and OLS slopes of the forward (inverse) relation with $\alpha^\perp$ and $\alpha^\parallel$ $(\beta^\perp$ and $\beta^\parallel)$, respectively. The intrinsic scatters are calculated via defining a projection vector from the relevant slopes $(\alpha^\perp, \alpha^\parallel, \beta^\perp, \beta^\parallel)$, as $\bm{P}=(1, -\text{slope})$. This way, we calculate the parallel intrinsic scatter from $\sigma^\parallel = \sqrt{\bm{P\Sigma^\prime P}^\intercal}$, where $\bm{\Sigma^\prime}$ is the relevant covariance matrix; $\bm{\Sigma^\prime}=\bm{\Tilde{\Sigma}}$ or $\bm{\Sigma^\prime}=\bm{\Sigma_{aa}}$. Then, orthogonal intrinsic scatter can be obtained with, e.g.,  $\sigma^\perp = \sqrt{\bm{P\Sigma^\prime P}^\intercal} / \sqrt{1 + {\alpha^\perp}^2}$ where $\bm{P} = (1, -\alpha^\perp)$.

In principle, in the absence of gross systematic errors in the measurements, $M_\star$ should strictly be less than $M_\mathrm{dyn}$. While this could be enforced by some choice of prior, in order to fairly test this expectation, we choose not to implement this prior. For instance, \cite{taylor2010} have seen that simple dynamical masses calculated under homology assumption ($k=$ constant) have resulted in $M_\star>M_\mathrm{dyn}$ for $M_\star<10^{10.5} M_\odot$, however this inconsistency has been removed when dynamical masses were calculated with $k(n)$, which suggests a systematic error in the values of $M_\mathrm{dyn}$ estimated using a constant $k$. Alternatively, we can use the expected (soft) limiting constraint due to, e.g., the choice of IMF. As a simple example, adding 0.3 dex to the $M_\star$ estimates to approximate the change of IMF from a \citet{chabrier2003} to a \citet{salpeter1955} IMF violate the $M_\star < M_\mathrm{dyn}$ expectation. Because we see that our fits in any case satisfy this expectation, we do not see the need to enforce a $M_\star > M_\mathrm{dyn}$ prior. (But see \ref{sec:unmodelled_random_errors} for some additional nuance around this point.)

\begin{figure*}
    \centering
    \includegraphics[width=\textwidth]{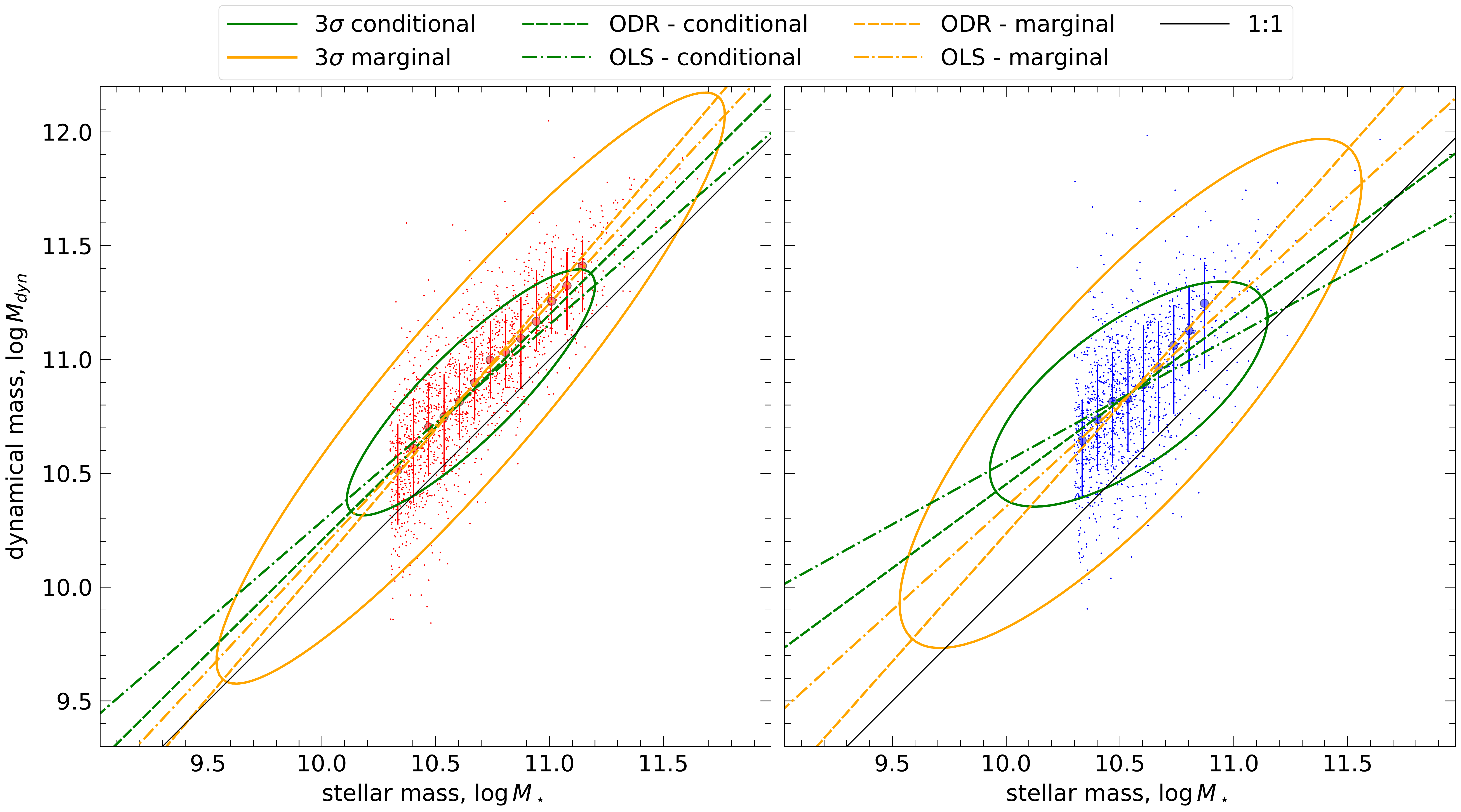}
    \caption{The relation between stellar mass and dynamical mass for quiescent (left) and star-forming galaxies (right). The red/blue dots show the data as observed. Large red/blue circles show the median, and the error bars show the 16\textsuperscript{th} and 84\textsuperscript{th} percentiles of $\log\,M_\text{dyn}$, in bins of $\log M_\star$. Solid green and orange ellipses show the underlying $3\sigma$ Gaussian distributions which are derived using the conditional and marginal distributions respectively, of $(\log\,M_\star, \log\,M_\text{dyn})$ pair obtained from the 8D model. Dashed lines that pass through the long axis of each ellipse are the best-fitting linear relations calculated from the eigen-decomposition of the corresponding covariance matrices, i.e., ODR with slopes $\alpha^\perp_\text{cond}$ and $\alpha^\perp_\text{marg}$, with green and orange representing conditional and marginal distributions respectively. Dash-dotted lines show the OLS linear relation with slopes $\alpha^\parallel_\text{cond}$ and $\alpha^\parallel_\text{marg}$. The solid black line shows the one-to-one relation.}
    \label{fig:mstar_mdyn_2d}
\end{figure*}

In Figure \ref{fig:mstar_mdyn_2d}, we compare the values of $\log M_\star$ and $\log M_\text{dyn}$ for both types of galaxies where the red/blue dots show the data for Qs/SFs. Here we give the inverse relation: $M_\text{dyn}$ as a function of $M_\star$. In each panel, the large circles show the median and their associated error bars show the 16/84 percentiles of $M_\text{dyn}$ in bins of $M_\star$. The solid ellipses in each panel show the $3\sigma$ Gaussian ellipse derived from the marginal in orange and for the conditional distribution in green. Dashed and dot-dashed lines show the best-fitting ODR and OLS relations, respectively, for both distributions. The black lines show the one-to-one relation. Finally, in Table \ref{tab:cond_marg}, we give the values of the slopes, intrinsic scatters and the correlation coefficients derived from both conditional and marginal distributions, for both forward and inverse fits, and for both galaxy populations. 

\begin{table}
       \centering
       \caption{The relation between $M_\star$ and $M_\text{dyn}$ for both galaxy populations inspected in different cases: Forward $(M_\star \propto M_\text{dyn}^\alpha)$ and inverse $(M_\text{dyn} \propto M_\star^\beta)$ fits derived from conditional and marginal distributions. We give both ODR and OLS slopes and the corresponding intrinsic scatters. We also give the correlation coefficients, only for the forward fits, because they will be the same for the inverse relations. This table unambiguously shows that, regardless of galaxy type, for both empirical prediction and astrophysical interpretation purposes, the ideal $M_\star-M_\text{dyn}$ relation with high correlation and low intrinsic scatter is achieved in conditional distributions: at fixed velocity dispersion, S\'ersic index and colour.}
       \resizebox{0.47\textwidth}{!}{
     \begin{tabular}{crccrcc}
\cmidrule{3-7}           &       & \multicolumn{2}{c}{$M_\star\propto M_\text{dyn}^\alpha$} &       & \multicolumn{2}{c}{$M_\text{dyn}\propto M_\star^\beta$} \\
\cmidrule{3-7}           &       & Quiescent & Star-forming &       & Quiescent & Star-forming \\
     \midrule
     \multirow{5}[2]{*}{Conditional} & $\alpha^\perp$ & $1.008\pm0.009$ & $1.358\pm0.027$ & $\beta^\perp$ & $0.992\pm0.009$ & $0.737\pm0.015$ \\
           & $\alpha^\parallel$ & $0.877\pm0.007$ & $0.839\pm0.016$ & $\beta^\parallel$ & $0.865\pm0.007$ & $0.551\pm0.011$ \\
           & $\rho$ & $0.871\pm0.003$ & $0.68\pm0.009$ & -     & -     & - \\
           & $\sigma_\alpha^\perp$ & $0.077\pm0.001$ & $0.101\pm0.001$ & $\sigma_\beta^\perp$ & $0.077\pm0.001$ & $0.101\pm0.001$ \\
           & $\sigma_\alpha^\parallel$ & $0.105\pm0.001$ & $0.147\pm0.002$ & $\sigma_\beta^\parallel$ & $0.105\pm0.001$ & $0.119\pm0.001$ \\
     \midrule
     \multirow{5}[2]{*}{Marginal} & $\alpha^\perp$ & $0.849\pm0.006$ & $0.889\pm0.013$ & $\beta^\perp$ & $1.178\pm0.008$ & $1.125\pm0.016$ \\
           & $\alpha^\parallel$ & $0.794\pm0.005$ & $0.748\pm0.011$ & $\beta^\parallel$ & $1.074\pm0.007$ & $0.909\pm0.012$ \\
           & $\rho$ & $0.923\pm0.002$ & $0.825\pm0.006$ & -     & -     & - \\
           & $\sigma_\alpha^\perp$ & $0.109\pm0.001$ & $0.147\pm0.002$ & $\sigma_\beta^\perp$ & $0.109\pm0.001$ & $0.147\pm0.002$ \\
           & $\sigma_\alpha^\parallel$ & $0.141\pm0.002$ & $0.189\pm0.003$ & $\sigma_\beta^\parallel$ & $0.164\pm0.001$ & $0.208\pm0.002$ \\
     \bottomrule
     \bottomrule
     \end{tabular}%
         }
       \label{tab:cond_marg}
\end{table}

Figure \ref{fig:mstar_mdyn_2d} and Table \ref{tab:cond_marg} show that both orthogonal and parallel (or direct) fits obtained from marginal distributions do not differ much between populations: they are generally consistent within their errors. Though, the fits obtained from conditional distributions differ considerably. Furthermore, correlation between $M_\star$ and $M_\text{dyn}$ is significantly smaller when fitted using conditionals. As expected, this shows that some part of the higher correlation encountered in the marginal fits are driven by the cross-correlations between $(\bm{m^*}, \bm{m^d})$ and $(\bm{s}, \bm{\nu}, \bm{c})$. Notably, tighter correlations exist in all cases for Qs.

Since the orthogonal slopes of forward and inverse relations are reciprocal of each other ($\alpha^\perp = 1/\beta^\perp$, which also leads to $\sigma_\alpha^\perp = \sigma_\beta^\perp$), as seen from Table \ref{tab:cond_marg}, the orthogonal slope of either relation for Qs is close to unity, albeit only when derived from conditional distributions.

The slopes derived from parallel fits $(\alpha^\parallel $ and $\beta^\parallel)$ to both forward and inverse relations are consistent with each other for Qs. The small parallel scatters for Qs $(\sigma_\alpha^\parallel$ and $\sigma_\beta^\parallel=0.105)$\footnote{As seen from Table \ref{tab:cond_marg}, in case of conditional distributions, $\sigma_\alpha^\parallel=\sigma_\beta^\parallel$ for Qs, because the slopes of the forward and inverse relations for Qs are almost equal $(\alpha^\parallel \approx \beta^\parallel)$ within $1\sigma$.} show that both $M_\star$ and $M_\text{dyn}$ can be predicted well from one another. However, in the case of SFs, it is harder to predict $M_\star$ from $M_\text{dyn}$ since the scatter in the direction of $M_\star$ reaches to 0.147 dex.

Our OLS-derived slope of the $M_\star$-$M_\mathrm{dyn}$ relation (marginal) for Qs is consistent with previous studies of \cite{taylor2010, cappellari2013, zahid2017}, who find a slope close to unity, with the implication that the central dark matter content is constant and/or a small fraction of the stellar content. The same is not true for SFs, where the ODR-derived slope is considerably smaller. Similarly, after performing the same analysis with the assumption of homology (see Appendix \ref{sec:homology}), we reach the same conclusion as \cite{taylor2010, zahid2017} in the aspect that this direct proportionality, $M_\star \propto M_\text{dyn}$, is possible only if (radial) non-homology is taken into account, at least for Qs. Interestingly, the orthogonal slopes in Table \ref{tab:cond_marg_k5} differ from unity by only $\sim 3\sigma$, suggesting that the effect of radial non-homology is still weak. However, this does not seem to apply to SFs, even when radial non-homology is accounted for by $k=k(n)$: we find that regardless of homology, the slope of the relation between $M_\star$ and $M_\text{dyn}$ for given $\sigma_e$, $n$ and $(g-i)_\text{rest}$, largely differs from unity (see, Tables \ref{tab:cond_marg} and \ref{tab:cond_marg_k5}).

That said, we emphasise that OLS is not the right method to infer the intrinsic, underlying correlation. Instead, we need to look at the ODR, which shows that the slopes of the $M_\star$--$M_\mathrm{dyn}$ relation for \textit{both Qs and SFs} are 1.)\ substantially differ from unity, and 2.)\ broadly comparable between the two populations: $\alpha = 0.849 \pm 0.006$ and $\alpha = 0.889 \pm 0.13$ for Q and SF galaxies, respectively. These imply that $f_\mathrm{DM}$ varies as a function of mass in the inner parts of both Q and SF galaxies.

Next, we check the effects of using the structure correction factor of \cite{cappellari2006} in equation (\ref{eq:bn}), which has been derived empirically, thus potentially provides a better prior on structural differences as a function of S\'ersic index. Replacing $\bm{m^d}$ calculated using equation (\ref{eq:kvn}), with $\bm{m^d}$ calculated via equation (\ref{eq:bn}), and then repeating the analysis, we find that the results given in Table \ref{tab:cond_marg} does not undergo any significant changes as seen in Table \ref{tab:kn_cappellari} that we give in Appendix \ref{sec:homology}. Though, the choice of $k(n)$ is expected to affect the dependence of $M_\star/M_\mathrm{dyn}$ relation on other galaxy properties.

Of course, we could have just fitted the $M_\star-M_\text{dyn}$ relation from the very beginning. The advantage of this approach is that it allows us to conveniently account for all the interrelations within the parameter space and makes possible to disentangle the trends stemming from other parameters, which is the topic of section \ref{sec:trends_mstar_to_mdyn}.

\section{Results II: Stellar-to-dynamical mass ratio}\label{sec:trends_mstar_to_mdyn}

This framework enables us to easily tackle questions such as, what parameters correlate with the stellar-to-dynamical mass ratio ($\log\,M_\star/M_\text{dyn}\equiv \bm{m^*_d}$ for brevity) at fixed, e.g., $M_\star$? In other words, we can isolate the effect of any one parameter on the $M_\star/M_\text{dyn}$ ratio when everything else is fixed. We will name this as \textit{isolated trends} for convenience (similar to partial correlations and their contributions to the overall correlation in partial linear regression, as noted in section \ref{sec:marg_vs_cond}).

If $y$ can be expressed as a linear combination $y=\sum_i a_i x_i$, the isolated trend of $x_i$ for a given $i$ can be defined as $\Delta_{x_i}(y)\equiv (\partial y/\partial x_i) x_i$. Extending this definition for $\bm{m^*_d}$, the isolated trend for each $\bm{x}\in \bm{Y_b} = (\bm{m^*}, \bm{s}, \bm{\nu}, \bm{c})$ becomes $(\partial \bm{m^*_d} / \partial \bm{x}) \bm{x}$. The slope, $\partial \bm{m^*_d} / \partial \bm{x}$, of each trend can be calculated via equation~(\ref{eq:tilda}) for the conditional distribution $(\bm{m^*_d}, \bm{x}\,|\,\bm{Y_b \setminus \{x\}})$. 

\begin{figure*}
    \centering
    \includegraphics[width=\textwidth]{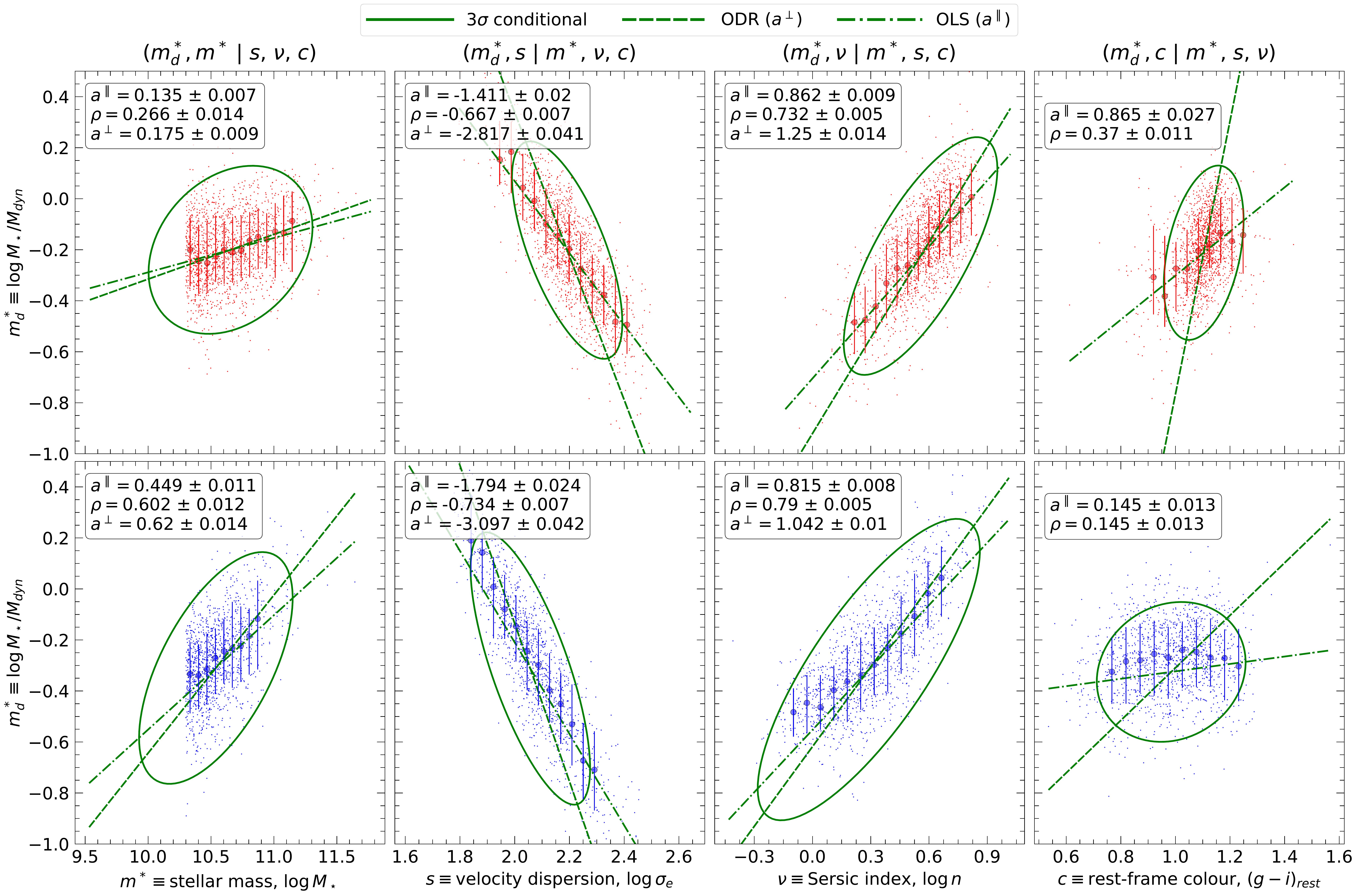}
    \caption{Isolated trends of $M_\star/M_\text{dyn}$ ratio calculated from the conditional distributions given on top of each column. Colours and symbols are the same as Figure \ref{fig:mstar_mdyn_2d}. OLS slopes $(a^\parallel)$ and the \textit{true} correlation coefficients $(\rho)$ are given in the upper middle part of each panel. ODR slopes $(a^\perp)$ are not given for $(g-i)_\text{rest}$ since they are not meaningful when $m^*_d$ is plotted as a function of colour, as seen here. The constrained layout of this figure prevents setting one-to-one aspect ratio, thus, the lines from ODR fits do not align with the major axes of the corresponding ellipses.}
    \label{fig:msmd_conditional}
\end{figure*}

Figure \ref{fig:msmd_conditional} shows these isolated trends of $\bm{m^*_d}$ for Qs in upper panels and for SFs in lower panels. In addition to the data points, each panel is comprised of the $3\sigma$ Gaussian ellipse, ODR and OLS best-fit lines, all derived from the conditional distribution. In the upper left part of each panel, we give the OLS and ODR slopes of the relevant isolated trend and the correlation coefficient between $\bm{m^*_d}$ and the parameter in the horizontal axis.

The results presented in Figure \ref{fig:msmd_conditional} show that for both populations, $M_\star/M_\text{dyn}$ ratio is a strong function of $\sigma_e$ at fixed $M_\star, n, (g-i)$; and S\'ersic index at fixed $M_\star, \sigma_e, (g-i)$ with similar variations between populations. At fixed $M_\star, \sigma_e$ and $n$, there is a strong variation with $(g-i)$ for Qs, however, it is quite weak for SFs. Moreover, variation with $M_\star$ is substantially strong for SFs while it is weak for Qs. Notably, the strong variations with $\sigma_e$ and $n$ seen here pose an apparent contradiction to the results of \cite{taylor2010} who found that when non-homology was taken into account with $k(n)$ in equation~(\ref{eq:kvn}), the trends of $m^*_d$ with central velocity dispersion $(\sigma_0)$ and $n$ were small. They found the slopes for $\log\,\sigma_0$ and $n$ to be -0.18 and 0.01, respectively, which are significantly smaller than our results in Figure \ref{fig:msmd_conditional}. This is because the results presented in Figure 5 of \cite{taylor2010} are of 2D-fitting applied to the 2D projections of the data they have used: pairs of $m^*_d-m^*$, $m^*_d-n$ and $m^*_d-s$. With a more rigorous analysis, here we can isolate the correlations between $m^*_d$ and both $s$ and $\nu$, finding strong correlations for both, but in opposing senses. Because $s$ and $\nu$ are themselves correlated, the simplistic analysis of \cite{taylor2011} was unable to disentangle these confounding effects, and instead saw flat residuals in $m^*_d$ when plotted as a function of either $s$ or $\nu$.  Nevertheless, they reach results similar to ours based on the observed scatter in $m^*_d$ in bins of $n$ and $m^*$ (their Figure 6), which is another way of showing the actual correlations at fixed $n$ and $m^*$. They find that the scatter in $m^*_d$ strongly varies with $n$ at fixed $m^*$ and vice versa.

Finally, in Figure \ref{fig:msmd_trends_cappellari}, we show these trends when equation (\ref{eq:bn}) is adopted. Although the choice between equation (\ref{eq:kvn}) and (\ref{eq:bn}) does not seem to create a remarkable difference in the relation between $M_\star$ and $M_\mathrm{dyn}$, we see that it certainly affects the interpretation of the intrinsic correlation between $M_\star/M_\mathrm{dyn}$ ratio and $n$. As seen in Figure \ref{fig:msmd_trends_cappellari}, while almost all the best-fit parameters $(a^\parallel, a^\perp, \rho)$ for the trends with $M_\star, \sigma_e$ and $(g-i)$ colour agree within errors when compared to Figure \ref{fig:msmd_conditional}, the correlation with $n$ drastically decreases for both populations when $k(n)$ in equation (\ref{eq:bn}) is used.

\section{Results III: Exploring potential systematic errors in stellar mass estimates}\label{sec:mstar_precision}

The overarching goal in this paper is to evaluate the consistency between stellar and dynamical mass estimates, with a view to evaluating the potential for systematic errors in the stellar mass estimates.

Our method for tackling this problem can be better understood if we further deconstruct $M_\star/M_\mathrm{dyn}$ ratio into its constituents, in a similar way as in \cite{graves2010_3},
\begin{equation}
    \frac{M_\star}{M_\mathrm{dyn}} = \left[ \frac{M_{\star, \mathrm{IMF}}}{L} \frac{M_\star}{M_{\star, \mathrm{IMF}}} \frac{M_\mathrm{tot}}{M_\star} \frac{M_\mathrm{dyn}}{M_\mathrm{tot}} \frac{L}{M_\mathrm{dyn}}\right] \frac{M_\star}{M_\mathrm{dyn}}~.
    \label{eq:mdyn_breakdown}
\end{equation}
Here the first term, $M_{\star, \text{IMF}}/L$, is the stellar-mass-to-light ratio measured through SPS by assuming a fiducial IMF, whereas $M_\star$ is the true stellar mass (i.e., with the true IMF), therefore the second term, $M_\star/M_{\star,\text{IMF}}$, encapsulates the differences in the IMF. Because the total mass $(M_\mathrm{tot})$ comprises of baryonic mass and dark matter mass; $M_\mathrm{tot} = M_\mathrm{bar} + M_\mathrm{DM}$, the term $M_\star/M_\mathrm{tot}$ encapsulates both non-stellar baryons (i.e. gas, dust, etc.), as well as the dark matter fraction, $f_\mathrm{DM} = M_\mathrm{DM}/M_\mathrm{tot}$. Thus, we can separate the uncertainties in SPS-modelling from the ones in $k(n)$, i.e., the first three terms in square brackets can be attributed to $f_\star$ of equation (\ref{eq:f_mstar_f_mdyn}) and the last two terms to $f_\mathrm{dyn}$. However, as seen in the term $(M_\star/M_{\star, \mathrm{IMF}})(M_\mathrm{tot}/M_\star)$, the possible variations in the IMF and in the dark matter fraction are indistinguishable from each other. This remains true even with spatially resolved spectroscopy. 
Nevertheless, using the simple $M_\mathrm{dyn}$ estimate as an SED-independent check on $M_\star$ provides a way to see the residual trends as a function of SPS-modelling constituents, but the astrophysical interpretation of these trends is still limited by the degeneracy between IMF and DM.

\subsection{Calibrating a dynamical estimator for stellar mass}\label{sec:mstar_dyn}

To address the difficulties outlined in section \ref{sec:motivation}, our approach is to define a stellar mass proxy $(\hat{M}_\star)$. In the light of our results summarised in Figure \ref{fig:msmd_conditional}, which shows that $M_\star/M_\text{dyn} \sim M_\star/\sigma_e^2 R_e$ ratio has strong variations with $\sigma_e, n$ and $(g-i)_\text{rest}$, we can write $\hat{M}_\star$ as a \textit{hyperplane} given by,
\begin{align}
    \log \hat{M}_\star = a_0 \log(\sigma_e^2 R_e) &+ a_1\log\sigma_e + a_2\log n \nonumber \\
    &+ a_3(g-i)_\text{rest} + a_4
    \label{eq:mstar_hyp}
\end{align}
which we then take as an SED-independent predictor for the stellar mass, and we fit for the $a_i$ coefficients.

From this definition of $\hat{M}_\star$, one way to understand this approach is thinking of it as fitting $M_\star$ vs $M_\text{dyn}$ with $\log k(n) \sim a_2 \log n$. When compared to equations (\ref{eq:f_mstar_f_mdyn}) and (\ref{eq:mdyn_breakdown}), this can also be interpreted as $\log[(f_\star/f_\mathrm{dyn})k(n)] \sim a_2 [\log (f_\star/f_\mathrm{dyn}) + \log n]$ where the correction factors $f_\star$ and $f_\mathrm{dyn}$ are eventually absorbed in the zero-point, $a_4$.
In this sense, by fitting for the values of $a_i$ we can calibrate a prescription for $M_\text{dyn}$ as the one that best predicts $M_\star$, and then other coefficients can be understood as describing variation in $M_\star/M_\text{dyn}$ as a function of velocity dispersion, structure, and colour.

We set the values of the coefficients by extracting the best-fit relation between the SED-derived stellar mass estimates, $M_\star$, and this proxy variable, $\hat{M}_\star$. Since our interest is in predicting the value of $M_\star$, we use the forward linear OLS regression (see, section \ref{sec:mstar_vs_mdyn}), derived by applying the linear transformation $\bm{\omega}: \mathbf{\hat{y}} \rightarrow \bm{Y^\prime}=(\bm{m}^*, \bm{2s+r}, \bm{s}, \bm{\nu}, \bm{c})$ to our parent model, as shown in section \ref{sec:lintransform}, then calculating the coefficients from the covariance matrix of $\bm{Y^\prime}$. Note that, this is equivalent to calculating the coefficient of each $\bm{x}$ in the right hand side of equation~(\ref{eq:mstar_hyp}) from the conditional distribution $(\bm{m^*}, \bm{x}\,|\,\bm{Y^\prime\setminus \{m^*, x\}})$, i.e., OLS regression of $(\bm{m^*}, \bm{x})$ at fixed everything else. 

Here we take an extra step and fit our model (equation \ref{eq:wbc_logsumexp}) separately to another 8D parameter space constructed by just replacing the stellar masses from \cite{taylor2011} with the ones from {\sc ProSpect} of \cite{robotham2020}. The $a_i$ coefficients and the correlation between $M_\star$ and $\hat{M}_\star$ acquired for both data sets are given in Table \ref{tab:mstar_hyp_coefficients}.

\begin{table}
   \centering
   \caption{Best-fitting coefficients $(a_i)$ of the stellar mass proxy $(\log\,\hat{M}_\star)$, scatters (observed, intrinsic and scatter due to measurement errors) and correlation between $\hat{M}_\star$ and $M_\star$ obtained for both methods of SED-fitting; simple and {\sc ProSpect}.}
       \resizebox{0.47\textwidth}{!}{
          \begin{tabular}{rcccc}
        \cmidrule{2-5}    & \multicolumn{2}{c}{$M_\star$ (Simple SED)} & \multicolumn{2}{c}{$M_\star$ ({\sc ProSpect} SED)} \\
        \cmidrule{2-5}    & Quiescent & Star-forming & Quiescent & Star-forming \\
          \midrule
          $a_0$ & 0.846$\pm$0.007 & 0.809$\pm$0.014 & 0.757$\pm$0.016 & 0.578$\pm$0.027 \\
          $a_1$ & -0.568$\pm$0.028 & -0.281$\pm$0.044 & -0.399$\pm$0.063 & -0.097$\pm$0.089 \\
          $a_2$ & 0.003$\pm$0.011 & 0.196$\pm$0.013 & -0.019$\pm$0.025 & 0.106$\pm$0.025 \\
          $a_3$ & 0.728$\pm$0.031 & -0.097$\pm$0.017 & 0.479$\pm$0.069 & 0.249$\pm$0.037 \\
          $a_4$ & 6.968$\pm$0.034 & 7.303$\pm$0.052 & 7.546$\pm$0.078 & 7.925$\pm$0.1 \\
          $\sigma_\text{obs}$ & 0.176 & 0.209 & 0.136 & 0.155 \\
          $\sigma_\text{int}$ & 0.108 & 0.147 & 0.067 & 0.083 \\
          $\sigma_\text{err}$ & 0.139 & 0.149 & 0.118 & 0.131 \\
          $\rho$ & 0.956 & 0.898 & 0.906 & 0.832 \\
          \bottomrule
          \end{tabular}%
     }
   \label{tab:mstar_hyp_coefficients}
\end{table}
 
For quantifying the mismatch between $M_\star$ and $\hat{M}_\star$, we define a mismatch parameter, $\delta_{m^*}$, as the intrinsic scatter of $\log\,M_\star/\hat{M}_\star$. As in section \ref{sec:mstar_vs_mdyn}, this can be calculated via constructing a projection vector from the best-fitting hyperplane coefficients as $\bm{P}=(1, -a_0, -a_1, -a_2, -a_3)$. The total observed scatter $\sigma_\text{obs}$ can be evaluated as the rms of $\sqrt{\bm{P} (\bm{\Sigma}_{m^*}+\mathbf{E_{j,m^*}}) \bm{P}^\intercal}$ where $\bm{\Sigma}_{m^*}+\mathbf{E_{j,m^*}}$ is the convolution of covariance and error matrices of parameters included in equation~(\ref{eq:mstar_hyp}) and can be obtained by applying the Jacobian of the transformation $\omega$ to the original convolution matrix $\bm{\Sigma} + \mathbf{E_j}$. Since observed scatter is the square root of the quadrature sum of intrinsic scatter and scatter due to measurement errors, $\sigma_\text{obs}^2 = \sigma_\text{int}^2 + \sigma_\text{err}^2$, the intrinsic is $\sqrt{\bm{P} \bm{\Sigma}_{m^*} \bm{P}^\intercal}$ and scatter from errors can thus be calculated from the rms of $\sqrt{\bm{P}\mathbf{E_{j,m^*}} \bm{P}^\intercal}$. The values for $\sigma_\text{obs}, \sigma_\text{int}$ and $\sigma_\text{err}$ are also given in Table \ref{tab:mstar_hyp_coefficients}.

\begin{figure*}
    \centering
    \includegraphics[width=\textwidth]{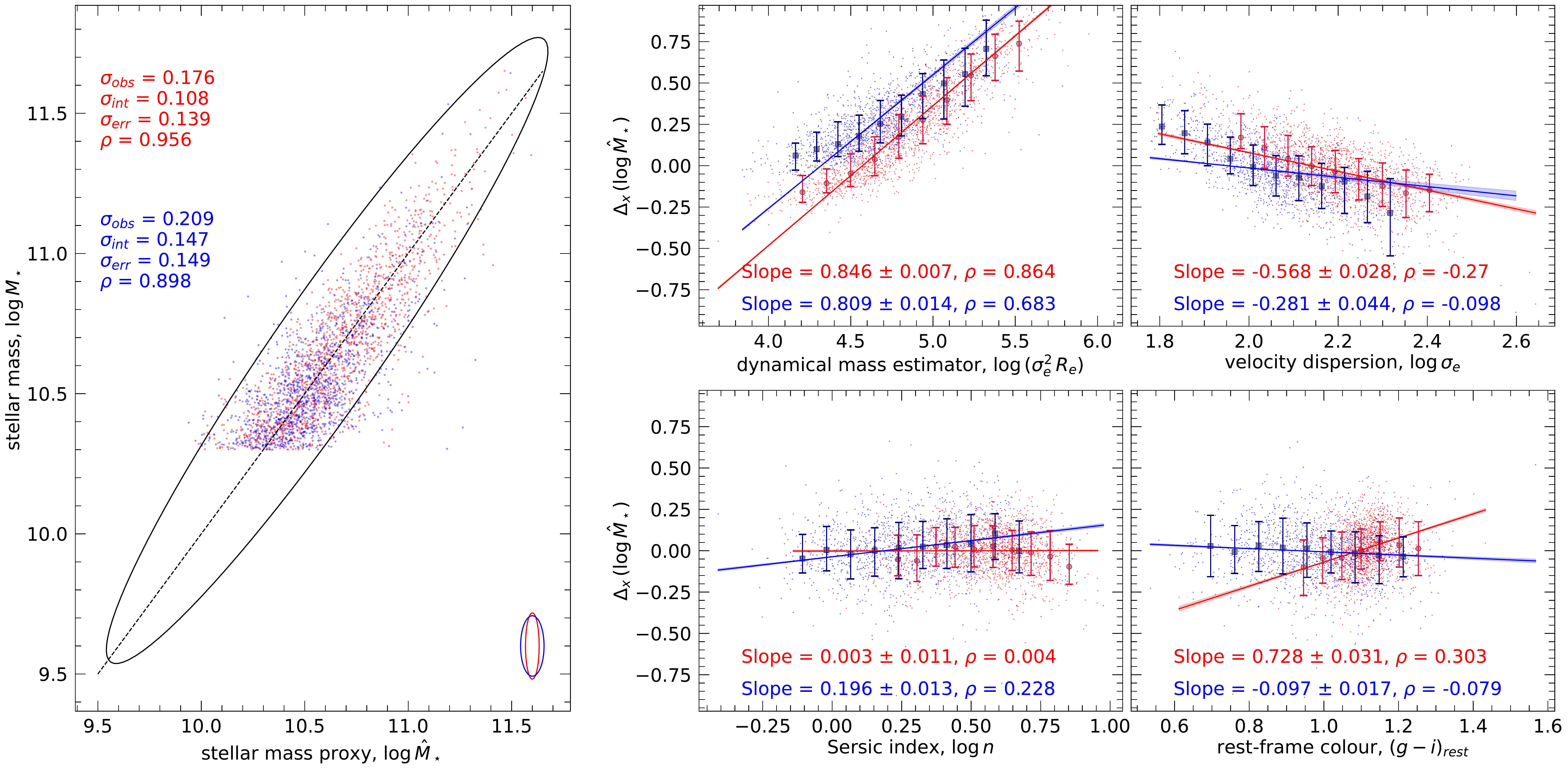}
    \caption{Calibrating the dynamical estimator of stellar mass (i.e., the stellar mass proxy) as a hyperplane for both galaxy populations. Blue and red colours represent the star-forming and quiescent galaxies respectively. Points show the data. \textit{Left panel:} Comparison of SED-derived stellar masses $(\log M_\star)$ to our stellar mass proxy $(\log \hat{M}_\star)$ given with the corresponding $3\sigma$ Gaussian ellipse in solid black. Dotted line represents one-to-one relation, $\rho$ is the correlation between $\log M_\star$ and $\log \hat{M}_\star$, and $\sigma$'s reflect the scatters around the mean relation. Median error ellipses are given in the bottom right corner. \textit{Right panel:} Isolated trends, showing the slopes, $a_i$, of the best-fitting mass hyperplane. The large blue squares, red circles and error bars are the same as in Figure \ref{fig:mstar_mdyn_2d}. Shaded regions around the lines show the uncertainty in the relevant slope and $\rho$ denotes the correlation between $\log \hat{M}_\star$ and the parameter in the $x-$axis.}
    \label{fig:mstar_hyp}
\end{figure*}

The basic summary of our results for the Q and SF populations are given in Figure \ref{fig:mstar_hyp}. In the first panel (Figure \ref{fig:mstar_hyp}), we show $M_\star$ plotted against the proxy variable $\hat{M}_\star$ with the observed, intrinsic and remaining scatters listed. In each panel of Figure \ref{fig:mstar_hyp}, we show the isolated trend, $\Delta_x (y)\equiv (\partial y/\partial x) x$, which, in addition to equation~(\ref{eq:tilda}), can also be calculated as the remaining trend of the parameter in the $x-$axis after the ones from the other fitted parameters are subtracted from the quantity in the $y-$axis. As an example, the isolated trend of $\nu\equiv\log n$ is $\Delta_\nu (\log \hat{M}_\star) \equiv \log \hat{M}_* - \left[a_0\log(\sigma_e^2 R_e) + a_1\log\sigma_e + a_3(g-i)_\text{rest} + a_4\right] = a_2\log n$, which means that the slope of the trend shown in each panel of Figure \ref{fig:mstar_hyp} is just the slope of the relevant parameter in equation~(\ref{eq:mstar_hyp}). Finally, this figure also gives the correlation coefficients between $\log \hat{M}_\star$ and each parameter, which are calculated via, again, $(\bm{m^*}, \bm{x}\,|\,\bm{Y^\prime\setminus \{m^*, x\}})$ and then using equation~(\ref{eq:tilda}).

Figure \ref{fig:mstar_hyp} shows that the major deterministic factor is the dynamical mass estimator; $\sigma_e^2 R_e$, whereas, structure is only a weak factor for SFs and it essentially has no effect for Qs. This is in fact surprising since lack of dependence of $M_\star/\sigma_e^2 R_e$ on structure is not expected considering the results for $M_\star/M_\text{dyn}$ given in Figure \ref{fig:msmd_conditional}. As for colour, it is significant only for Qs despite their narrow range in $(g-i)$, as seen in Figure \ref{fig:fitcorner}. Besides, for both populations, the mass proxy $\hat{M}_\star$ does not strongly depend on $\bm{s}$ as the $M_\star/\sigma_e^2 R_e$ does (Figure \ref{fig:msmd_conditional}).

\subsection{Quantifying the systematics as a function of stellar populations}\label{sec:sed_systematics}

Now that we have calibrated an SED-independent predictor of mass which tightly correlates with the SED-derived stellar masses $(\rho\sim 0.9)$, we can say that almost all of the observed scatter $(\sigma_\text{obs})$ in this relation at fixed $\hat{M}_\star$ (i.e., at fixed $\sigma_e, R_e, n$ and $(g-i)_\text{rest}$) is due to the uncertainties in the SED-derived masses. By analysing the residuals as a function of stellar population parameters, we can interpret how much mismatch correlates with these parameters and take them as indicative of systematic errors in $M_\star$.

The way that we are going to do this is by first plotting the residuals against rest-frame colour $(g-i)_\text{rest}$, equivalent widths of H$\alpha$ and H$\delta$ lines, 4000\AA break strength $(D_n4000)$, dust extinction $[E(B-V)]$, specific star formation rate (sSFR $=$ SFR$/M_\star$), luminosity-weighted mean stellar age $[\log\langle t_\star \rangle_\text{LW}/Gyr]$ and metallicity $[\log\,Z_\star/Z_\odot]$. Notice that these are the SP parameters which are not included in our multi-dimensional fit, except for $(g-i)_\text{rest}$. We then proceed to fitting a linear relation to the residuals as $\log \left( M_\star / \hat{M}_\star \right) \equiv \Delta\log M_\star = Ax + B $ where $x$ represents each SP parameter. Therefore, the dispersion of $x$ across our sample, $\sigma_x$, will propagate through the scatter via $A\sigma_x$. Following the notion from \cite{taylor2020}, we will name this quantity as implied dispersion, $\sigma_{\text{implied},x} \equiv A\sigma_x$, and it can be regarded as a lower limit on the contribution from $x$ to the scatter. We use $\sigma_\text{implied}$ and $\rho$ as metrics to identify which parameters explain the most/least variation, assuming that these residual trends are only due to systematic errors in the SED-derived stellar masses.

Even though what follows from now might seem to be a simple task of fitting lines to the residual trends, providing estimates of $\sigma_{\text{implied},x}$ as accurate as possible relies on a rigorous fit that also accounts for the covariant uncertainties in the SPS parameters: measurement errors in stellar mass are strongly correlated to the ones in colour, luminosity-weighted age and dust extinction\footnote{The correlation coefficients between these measurement errors are provided in the stellar masses DMU (\textit{StellarMassesGKVv24}) \citep{taylor2011}. }. As such, we implement a 2D-Gaussian model in \textit{STAN}, which is essentially the 2D and simplified version of our 8D model (e.g., $M=2$ in equation~\ref{eq:mdgauss}), to fit $\Delta \log M_\star$ as a function of each SPS parameter separately. $A, \sigma_x$ and the correlation $(\rho)$ between the residuals and each $x$ can be obtained from the corresponding covariance matrix resulting from these 2D fits. 

\begin{figure*}
    \centering
    \includegraphics[width=\textwidth]{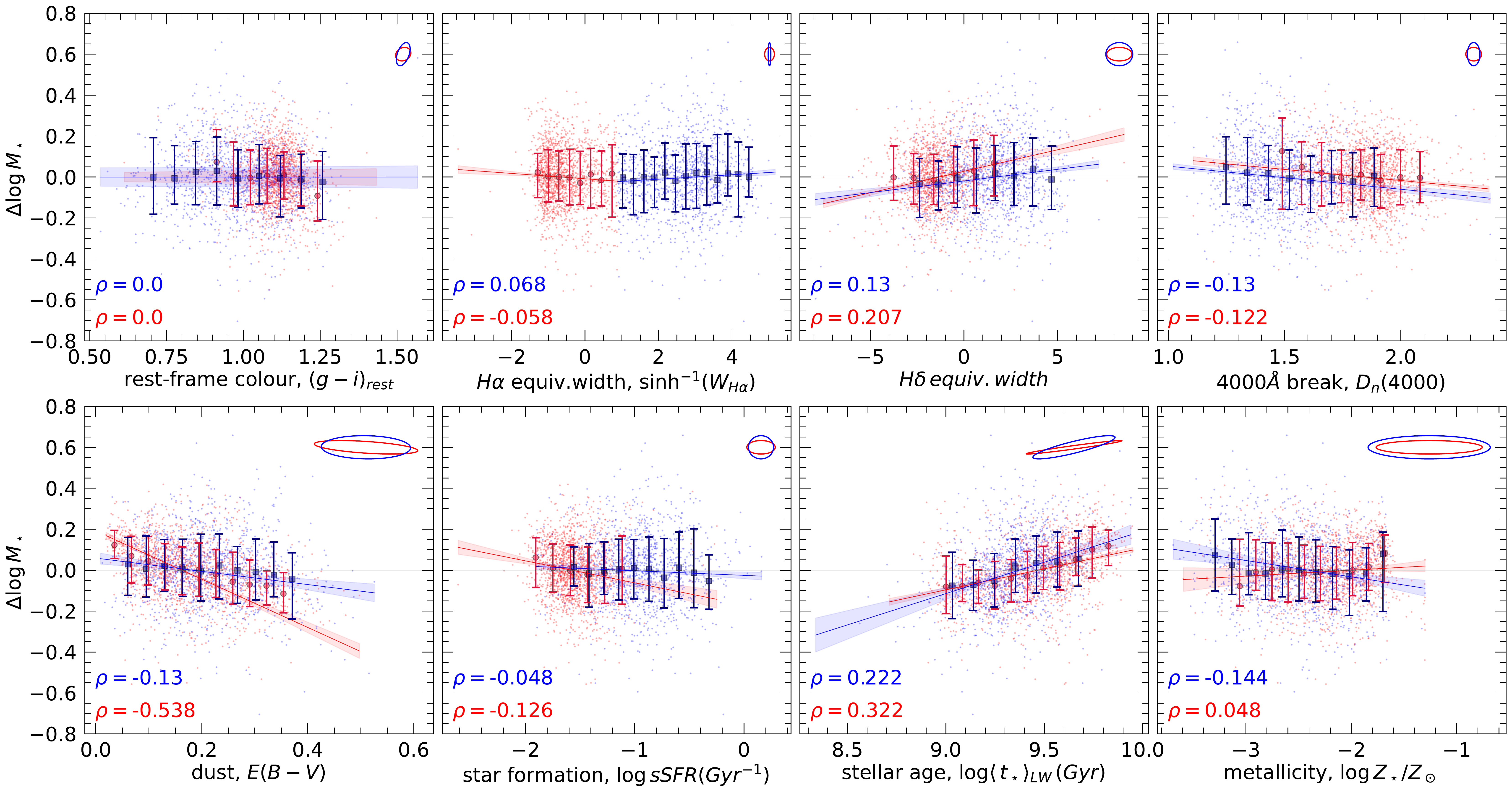}
    \caption{Residuals as a function of stellar population parameters, each fitted with a 2D-Gaussian model for both Q and SF galaxies. It should be noted here that the parameters in the top row are directly measured, while the ones in the bottom row are derived through SPS modelling. Correlation coefficients calculated from the model are given in the bottom left corner of each panel. Median error ellipses are shown at the top right corner. Symbols and colours are the same as Figure \ref{fig:mstar_hyp}.}
    \label{fig:mstar_sps_residuals}
\end{figure*}

The residuals plotted as a function of SP parameters are given in Figure \ref{fig:mstar_sps_residuals} which also shows the fitted lines. It should be noted here that even though some of the trends are in the same direction as the correlated errors, we can safely say that none of these trends are driven by the correlated errors since they have been accounted for in our 2D fitting algorithm as well.

The next question is what fraction of the mismatch between these mass estimates can be explained by contributions from the strong trends with SP parameters? Here, we quantify this by the fraction of the implied scatter from an SP parameter to the mismatch parameter, calculated in quadrature as $f_{x,\text{int} } = \sigma^2_{\text{implied}, x } / \delta_{m^*}^2$. The values inferred from our 2D-Gaussian models for $A, \sigma_x, \rho, \sigma_{x,\text{implied} }$ and finally $f_{x,\text{int} }$ are given in Appendix \ref{sec:more_discussion} (Table \ref{tab:spsparams}). 

\subsubsection{Potential systematics for star-forming galaxies}\label{sec:ltg_systematics}

It is interesting to see that the larger scatter in $M_\star$ for SF galaxies do not seem to be associated with the uncertainties in SP parameters. This might partly stem from that $M_\text{dyn}$, (thus, the simple dynamical mass estimator $\sigma_e^2 R_e$) cannot provide a prediction of $M_\star$ for SFs, as tight as it can for Qs, despite the strong correlation (see, Table \ref{tab:cond_marg}). 

Age, dust and metallicity together can account for $\sim 10\%$ of scatter, while star formation does not seem to play any role. It is difficult to draw definite conclusions because of the age-dust-metallicity degeneracy, since an increase in any of these parameters results in the overall reddening of the spectra. Additionally, the covariant errors in age-dust-metallicity introduce another complication, since the covariance values are not available in the data set and their calculation is not as simple as the ones in S\'ersic parameters that we show in Appendix \ref{sec:calc_coverrors}. Therefore, we can only say that the scatter correlates with them. We cannot conclude that the masses are wrong while age/dust/metallicity are right, or vice versa, and that the masses are wrong because of incorrect age/dust/metallicity. However, this should not be regarded as a possible inadequacy of our method to disentangle these aspects, while it is in fact an issue related to the shortcomings of the data, such as the covariance in uncertainties of these parameters.

It is also worth pointing out here that the correlation we found with the light-weighted SED-based ages is consistent with the independent age indicators $H_\delta$ and $D_n 4000$ (Figure \ref{fig:sps_gama}).

\subsubsection{Potential systematics for quiescent galaxies}\label{sec:etg_systematics}

As seen from Table \ref{tab:spsparams} in Appendix \ref{sec:more_discussion}, dust and age seem to be the dominant sources of mismatch for Qs, being potentially responsible together for $\sim 55\%$ of the mismatch, whereas, metallicity has no effect. Interestingly, the age and dust dependent systematics act in opposite senses for quiescent galaxies, in which younger stellar populations are more dust-extincted. However, as discussed in section \ref{sec:ltg_systematics}, we cannot make further interpretations due to the age-dust-metallicity degeneracy and their covariant errors. Nevertheless, as in SFs, the trend with age is consistent with H$\delta$ and $D_n 4000$.

\subsubsection{Projection effects}\label{sec:axis_ratio}

The projected axis ratio, denoted as $q=b/a$ where $a$ and $b$ are the semi-major and semi-minor axes respectively, has a significant effect on dynamical mass estimates as shown by \cite{wel2022}. Though, our 8D parameter space does not include $q$, neither does our stellar mass proxy in equation~(\ref{eq:mstar_hyp}). Therefore, it is reasonable to expect some residual trends with $q$, regardless of whether $M_\text{dyn}$ in the 8D parameter space is calculated with equation~(\ref{eq:kq}).

\begin{figure*}
    \centering
    \includegraphics[width=\textwidth]{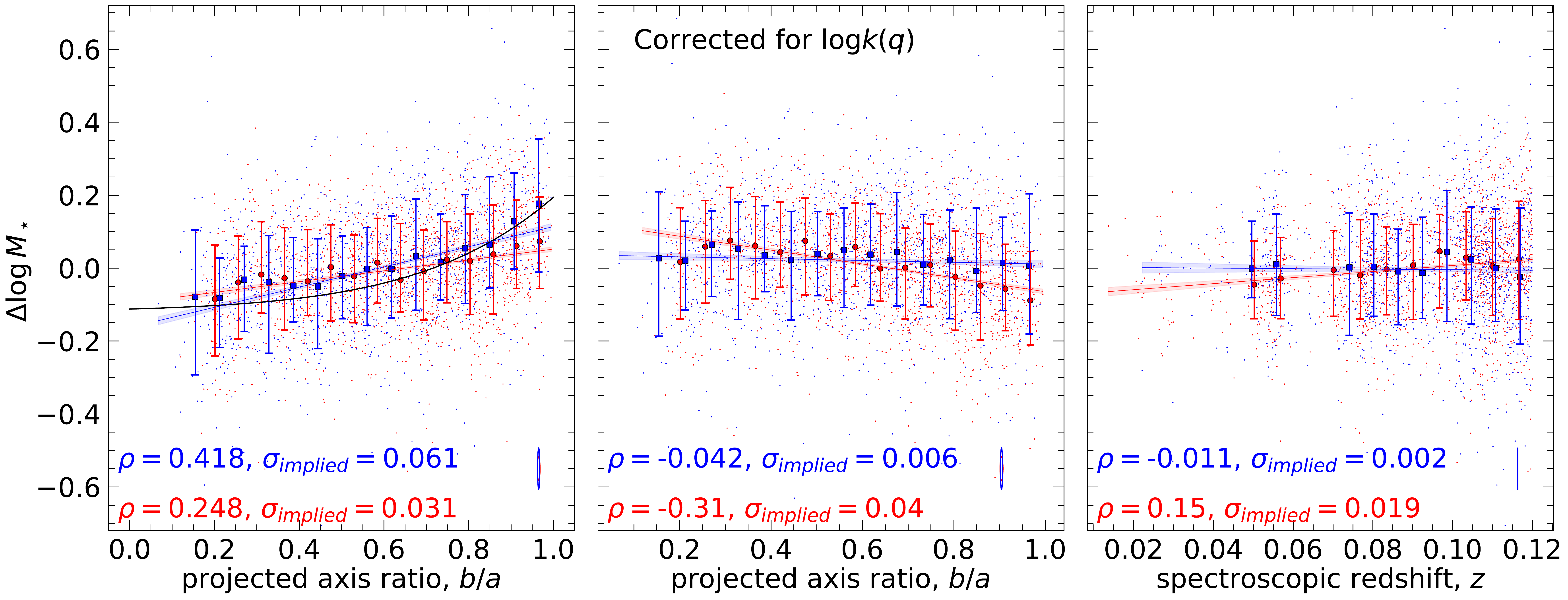}
    \caption{Residuals as a function of projected axis ratio and spectroscopic redshift. The bottom left corner of each panel gives the correlation and the scatter implied by the trend, whereas the bottom right corner shows the median uncertainties. The symbols and colours are the same as in Figure \ref{fig:mstar_sps_residuals}. The leftmost panel shows the residuals without correction for $\log k(q)$, which is the black curve. The middle panel shows the residuals with the correction: $\Delta\log M_\star - \log k(q)$.}
    \label{fig:axratio_z}
\end{figure*}

To address this concern, we apply the same procedure of fitting the residuals, this time as a function of projected axis ratio, obtained through S\'ersic fits, and also redshift. We give the results in Figure \ref{fig:axratio_z}, the leftmost panel of which shows that $q$ is a significant source of scatter in $M_\star$ for both Q and SF populations, if it is not accounted for. The effect is particularly pronounced for SFs, corresponding to $\sim18\%$ of the intrinsic scatter, which is larger than the total scatter that can be associated with SPS parameters. We also plot the $\log k(q)$ from equation~(\ref{eq:kq}) \citep{wel2022} with the smooth black curve in the leftmost panel, showing a good match to the trend. We then fit the residuals that are $\log k(q)$ subtracted in the middle panel. This correction successfully removes the trend for SFs, however, seems to over-correct the Qs and leaves a trend even slightly stronger in the opposite direction. We note that $k(q)$ correction has no effect whatsoever on our results so far. Finally, as seen in the rightmost panel, the residuals do not show notable variations with redshift for neither population.

\subsection{Potential unmodelled sources of random error: Constraints on IMF variability} \label{sec:unmodelled_random_errors}

After taking into account the significant scatter stemming from projection effects, the initial intrinsic scatters of 0.108 dex and 0.147 dex drop to 0.103 dex and 0.133 dex for Qs and SFs respectively.
From Figure \ref{fig:mstar_sps_residuals} and Table \ref{tab:spsparams}, we can see that SP parameters can explain $\sim 65\%$ of the intrinsic scatter for quiescent galaxies, however, they can account for only $\sim 13\%$ for star-forming galaxies. 
Under these circumstances, at a fixed set of SP parameters, the intrinsic dispersion between the SED-$M_\star$ and our dynamical estimates $\hat{M}_\star$ becomes 0.055 dex and 0.122 dex for Qs and SFs respectively. 

These numbers represent the combination of `true' variations and unmodelled sources of error in the measurements, therefore provide a soft upper limit on the unmodelled errors. The biggest caveat here is whether the errors in $M_\star$ due to the IMF are closely correlated to the errors in $M_\mathrm{dyn}$ due to the $f_\mathrm{DM}$. In that case, correlation between IMF and $f_\mathrm{DM}$ can cancel one another in $M_\star/M_\mathrm{dyn}$. Assuming that most of these unmodelled errors stem from the $M_\star$ measurements rather than $M_\mathrm{dyn}$, we can conclude that these values put a conservative limit on the ultimate precision in $M_\star$ estimates without, e.g., explicit IMF modelling for individual galaxies. By this argument, we come to a fundamental limit to precision of stellar mass estimates (in the face of IMF variations and any other stochastic factors) being better than 0.05 dex and 0.12 dex for Q and SF galaxies, respectively \citep[see also, e.g.][]{conroy2010a}.

Looking again at Figure 3 with these unmodelled sources of random error in mind, we make the following observations. While it is true that the main (i.e., the mean) relation satisfies the expectation that $M_\star < M_\mathrm{dyn}$, it is also true that there is a high $M_\star/M_\mathrm{dyn}$ tail to the population that seems to violate this constraint. This is most naturally explained by unmodelled errors in the stellar masses, dynamical masses, or both. We can subtract off our estimates for the unmodelled errors in the stellar masses, in which case the `intrinsic' scatters become 0.13 and 0.15 dex for Qs and SFs, respectively. 
This alleviates, but does not eliminate, the apparent violation of the $M_\star < M_\mathrm{dyn}$ expectation. 
Alternatively, we could have tried to enforce $M_\star < M_\mathrm{dyn}$ constraint by truncating the model (assuming that the mass estimates and their uncertainties are robust), and/or globally scale the mass estimates through changes to the IMF or $f_\mathrm{DM}$. At this point, however, we reach a point where we have exhausted what is possible to infer from single fibre spectroscopy: to go further will require proper dynamical modelling from spatially resolved spectroscopy (but still subject to the IMF/dark matter degeneracy).

\subsection{{\sc ProSpect} Masses}\label{sec:prospect}

We now apply the same analysis to the {\sc ProSpect} masses. This analysis is particularly valuable considering that there are not-so-negligible contributions to the scatter from metallicity and star formation rates, as seen in Table \ref{tab:spsparams} and Figure \ref{fig:mstar_sps_residuals}.

\begin{figure*}
    \centering
    \includegraphics[width=\textwidth]{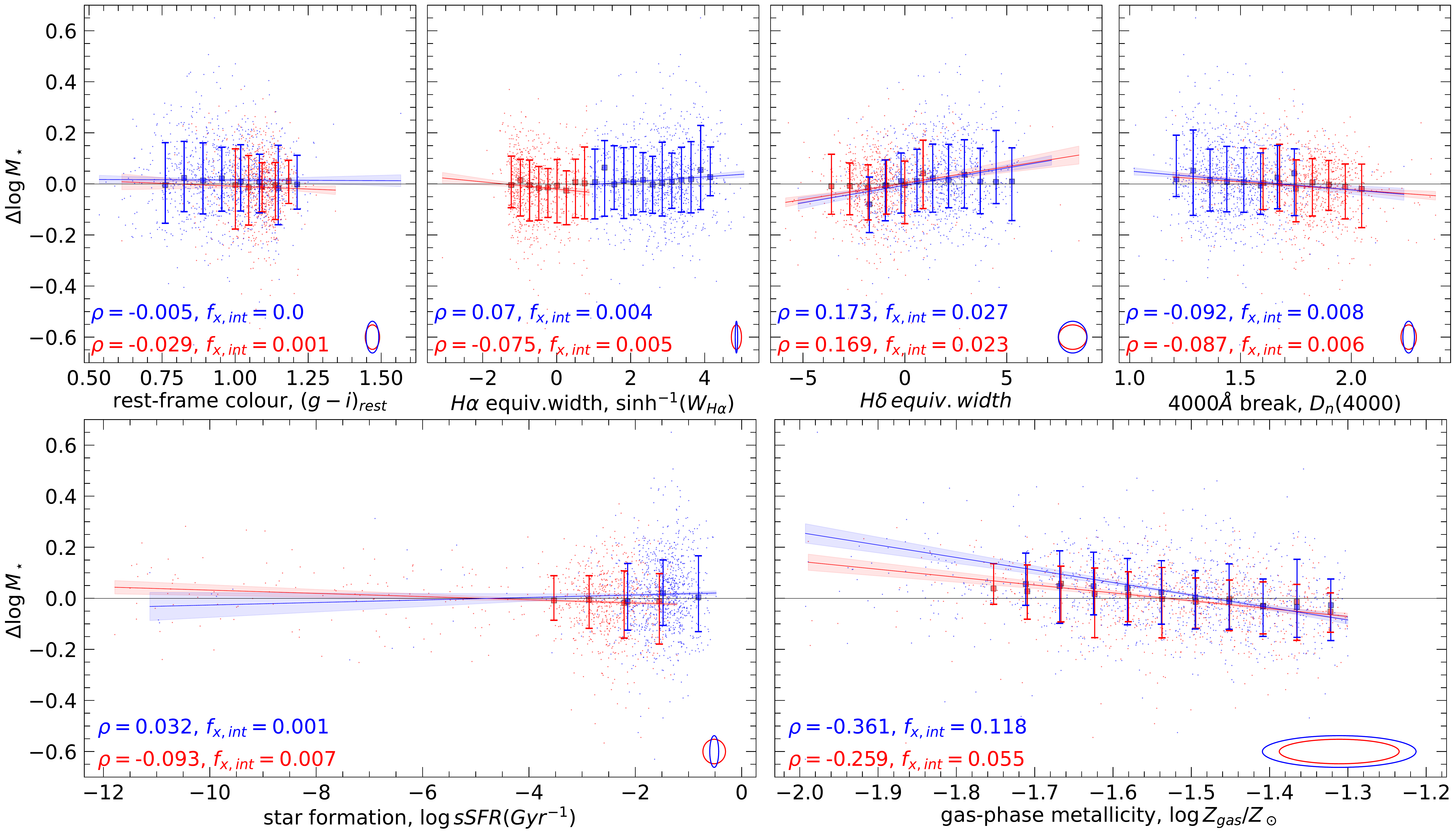}
    \caption{Same as Figure \ref{fig:mstar_sps_residuals}, excluding dust and mean stellar age, but for $M_\star$ estimates from {\sc ProSpect}.}
    \label{fig:mstar_sps_residuals_pro}
\end{figure*}

Figure \ref{fig:mstar_sps_residuals_pro} gives the residuals in $M_\star$ from {\sc ProSpect} as a function of SP parameters. Here, $s$SFR and $Z_\text{gas}$ are provided in {\sc ProSpect}, but dust and age are not. We also have to assume there is no covariance between the uncertainties of the parameters considered here, since this information is currently not available in {\sc ProSpect} data products.

When compared to Figure \ref{fig:mstar_sps_residuals}, we see that the apparent effect of SFR on scatter has vanished as expected because {\sc ProSpect} implements more complicated SFHs through time dependent $Z_\text{gas}$. However, we can see that there is a substantial contribution from $Z_\text{gas}$ to the mismatch, particularly in SFs, for which potential uncertainties in $Z_\text{gas}$ can explain $\sim 11\%$ of the scatter. Aside from $Z_\text{gas}$ and the minor contribution of H$\delta$, the SP parameters considered here do not seem to have any effect on the overall scatter. In this case, the remaining intrinsic scatter becomes 0.112 dex and 0.12 dex for Qs and SFs, respectively. Under the same aforementioned assumptions, this remaining scatter can be attributed to the joint effects of uncertainties in age, dust and IMF.

\section{Summary and conclusions}\label{sec:summary}

In this work, we have constructed a Bayesian framework that allows us to perform a multi-dimensional analysis of the GAMA data set, free from pernicious effects (to the extent of our knowledge) of selection biases and covariant errors that drive false and spurious trends. This allows us to perform a rigorous analysis of fundamental galaxy parameters while accounting for all the interrelations between them simultaneously. As a first application of our method, we analysed the consistency between stellar and dynamical masses. In the light of this analysis, we then examined the degree of precision of stellar mass estimates for GAMA galaxies in the local Universe by calibrating a stellar mass estimator from dynamical parameters; effective radius and velocity dispersion, which are independent of SED modelling. Even though a strong correlation was achieved between our fitted stellar masses and the SED derived ones, a significant scatter was still prevalent: $\sigma_\text{int}\sim 0.11$ dex for Qs and $\sim 0.15$ dex for SFs. Assuming that the scatter was due to the uncertain aspects of SED modelling, we then fitted the residuals as a function of SP parameters to quantify how much of the mismatch between these two stellar mass estimates could be attributed to them. Our main conclusions can be summarised as follows:

\begin{enumerate}
    \item Focusing first on OLS (i.e. direct/parallel fit) results: we reproduce results of past studies including \cite{taylor2010} and \cite{zahid2017}. For Qs, we find the best fitting slope of the $M_\star$--$M_\text{dyn}$ relation for given velocity dispersion, S\'ersic index and colour to be close to unity, which agrees with the results of \cite{taylor2010} and \cite{zahid2017}. Assuming homology (in this case, $k=5$), the slope differs from unity by only $\sim3\sigma$, suggesting that the effect of homology is weak for Qs. On the other hand, for SFs, regardless of the assumption of homology, the slope is substantially less than unity.
    \item Focusing instead on the ODR (i.e. perpendicular regression) results as the better means of inferring the true, underlying relation between stellar and dynamical mass, we see a rather different picture. Not only do both Q and SF populations show significant variation in $M_\star/M_\mathrm{dyn}$ as a function of mass, but the two populations show very similar variations: the inferred slopes are $\alpha = 0.849 \pm 0.006$ and $0.889 \pm 0.13$ for Q and SF galaxies, respectively.  The principal difference between the two populations is the larger scatter around the $M_\star$--$M_\mathrm{dyn}$ relation for SFs versus Qs: $\sigma = 0.189 \pm 0.003$ and $0.141 \pm 0.002$, respectively.
    \item The $M_\star/M_\text{dyn}$ ratio strongly varies with velocity dispersion and S\'ersic index for both galaxy populations. It also increases with increasing $M_\star$ for both populations, albeit much more weakly for Qs. Because the correlation of $M_\star/M_\mathrm{dyn}$ with $\sigma_e$ and $n$ are in opposite directions, they partially cancel each other out, which cannot be disentangled by considering only residuals from 1D fits like \cite{taylor2010}. This shows the need for and value of the multidimensional analysis of this paper.    
    \item Dust and age seem to be the largest contributor to the scatter for Qs, accounting for more than half of the scatter, while age and metallicity are major sources for SFs. The non-negligible effect of star formation in case of Qs indicate that some of the quiescent galaxies in the sample may have experienced recent starburst(s) which ended $\sim0.1-1$Gyr ago and are not accounted for by the smoothly declining single component star formation history adopted in SED-fitting. Though, this effect being little shows that this type of SFHs is sufficiently rigorous for SED-derived $M_\star$.
    \item Projected axis ratio is a significant source of uncertainty for both populations. The effect is larger than any one of SPS parameters in case of SFs. For future analysis, this could either be explicitly calibrated out in the Gaussian forward model; an alternative would be to use large IFS survey data sets to derive a better empirical prescription.
    \item The random scatter in SED-derived stellar masses of GAMA galaxies is small: 0.108 dex and 0.147 dex for Qs and SFs respectively, which can be decreased to 0.055 dex and 0.122 dex for Qs and SFs respectively, when the uncertain aspects of SPS models and projection effects are accounted for. 
    With simple assumptions, this value can be taken as a soft but conservative upper limit on the size of unmodelled sources of error in $M_\star/M_\mathrm{dyn}$, including e.g. IMF-related errors in the stellar masses, or variations in the gas and/or central dark matter content. While these different errors are inevitably degenerate, if we assume that this scatter is predominantly the result of IMF-related errors in the stellar mass estimates, this would imply a fundamental limit on the precision of stellar mass estimates of order $\sim 0.05$ dex and $\sim 0.12$ dex for Q and SF galaxies, respectively.
    \item Our results thus support the idea that $M_\star$ estimates are generally more reliable for Qs than for SFs \citep[][]{conroy2013} and that the potential impact of IMF variations is stronger for SFs \citep{conroy2010a}, but that stochastic (c.f. systematic) IMF variations do not produce errors greater than $\sim 0.1$ dex in the inferred stellar mass. The principal caveat to this statement is variation in the central dark matter content, $f_\mathrm{DM}$ and/or gas content. Unless the random scatter in $f_\mathrm{DM}$ is tightly correlated to IMF-related errors in the stellar mass, then our values represent conservative upper limits on unmodelled random errors in the stellar mass estimates; assuming some significant variation in $f_\mathrm{DM}$ would drive these limits down.  
    \item Conversely, under the assumption that the stellar masses are robust, this limit could be equally well interpreted in terms of the variation in $f_\mathrm{DM}$ (at fixed mass, S\'ersic index, velocity dispersion, colour, etc.). This interpretation would be broadly consistent with the observed scatter in the Tully-Fisher and Faber-Jackson relations, and with the implication that SF galaxies have a greater variation in $f_\mathrm{DM}$ than Q galaxies \citep[see, e.g.][]{posti2021} --- but with the important caveat that this interpretation hinges on the precision of the stellar mass estimates for SF versus Q galaxies.
\end{enumerate}

This framework provides important diagnostics for future galaxy census surveys like DESI-BGS \citep{hahn2022} and the 4MOST Hemisphere Survey \citep[4HS;][]{taylor2023}.

\section*{Acknowledgements}

We would like to thank the anonymous referee for providing helpful comments and suggestions crucial to significantly improving the flow of the not-so-straightforward concepts in this paper.
This work has been made possible by the \textit{STAN} \citep[][]{standevteam} software implemented in Python (PySTAN).
FDE acknowledges funding through the H2020 ERC Consolidator Grant 683184, the ERC Advanced grant 695671 ``QUENCH'' and support by the Science and Technology Facilities Council (STFC).

\appendix

\section{Calibrating Covariant Errors on S\'ersic-Fit Parameters}\label{sec:calc_coverrors}

\begin{figure*} 
    \centering
    \includegraphics[width=0.7\textwidth]{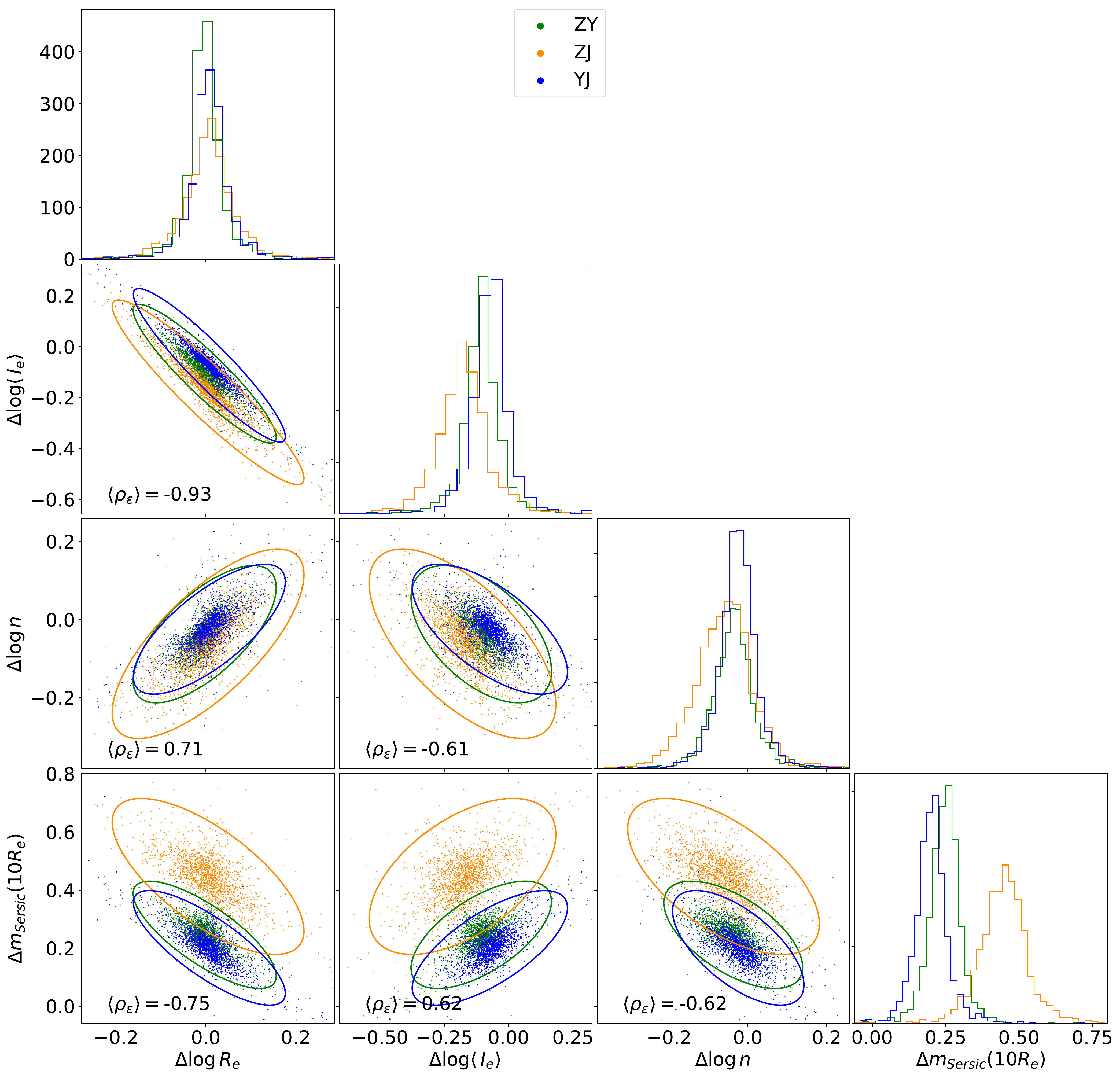}
    \caption{Distributions of variations in S\'ersic photometry parameters $r, i, \nu, m_\text{S\'ersic}(10R_e)$ for pairs of independent filters, i.e., as a function of wavelength, for quiescent galaxies. The off-diagonal panels show the bivariate distributions of the variations along with their 99\% confidence ellipses. The effect of interdependence between S\'ersic parameters is clearly seen from the strong correlations between these variations, which lead to correlated uncertainties. At the bottom left of each off-diagonal panel, the mean correlation coefficient for the uncertainties $\langle\rho_\varepsilon\rangle$ derived from the three filter pairs is also given.}
    \label{fig:covsersic}
\end{figure*}

One way to quantify the covariant errors in S\'ersic parameters -- effective radius, total magnitude and S\'ersic index -- is to examine the variations in these parameters as a function of wavelength against each other. To do this, we employ the method of \cite{magoulas2012} and obtain the distribution of differences in each S\'ersic parameter $(\Delta x)$ for pairs of independent VIKING-\textit{ZYJHK} filters: e.g., $\Delta x = x_Z-x_Y$. In Figure \ref{fig:covsersic}, for Qs, we present these distributions for \textit{ZY, ZJ, YJ} filter pairs, in which covariance of these variations is clearly seen, leading to covariant uncertainties in S\'ersic parameters. The correlation coefficients for uncertainties in each parameter pair $(x,y)$ are calculated from these covariances with,
\begin{equation}
    \rho^\varepsilon_{xy} = \frac{\text{Cov}(\Delta x, \Delta y)}{\sigma_{\Delta x}\sigma_{\Delta y}}
\end{equation}
individually for each filter pairs. Here, $\sigma_{\Delta x}$ and $\sigma_{\Delta y}$ are the standard deviations of $\Delta x$ and $\Delta y$. We can then use the mean of the values from all filter pairs in constructing the error matrices as in equation~(\ref{eq:fullerrormatrix}).

\section{Analytical background of the framework}\label{sec:analytical_gaussian}

In this appendix, we present the details on the mathematical and statistical foundations of our hierarchical Bayesian framework.

We adopt a common notation in which boldface letters refer to vectors and matrices. The main objective is to find the mean vector $(\bm{\mu} \text{ or } \mathbf{\bar{y}} \in\mathbb{R}^M)$ and the covariance matrix $(\bm{\Sigma}\in\mathbb{R}^{M\times M})$, which define the $M$-dimensional Gaussian, based on the observed data $\mathbf{\hat{y}}=(\mathbf{\hat{y}_1}, \dots, \mathbf{\hat{y}_M})\in\mathbb{R}^{N\times M}$ where each $\mathbf{\hat{y}_i}$ for $i=1,\dots,M$ represents a column vector of $N$ observations. Using a hierarchical approach, our goal is to find an expression for the posterior probability density of the parameters $(\theta)$ and hyperparameters $(\varphi)$ given the data: $p(\varphi, \theta | \mathbf{\hat{y}})$. 

Let $\theta=\mathbf{y}$ where $\mathbf{y}\in\mathbb{R}^{N\times M}$ is the true (latent) variables, and let $\varphi=\{ \bm{\mu}, \bm{\Sigma} \}$. Using Bayes' theorem, we can write
\begin{equation}
    p(\varphi, \theta | \mathbf{\hat{y}}) \propto p(\varphi)p(\theta | \varphi) p(\mathbf{\hat{y}}|\theta)
    \label{eq:joint_post}
\end{equation}
where $p(\varphi)$ is the hyperprior distribution, $p(\theta | \varphi)$ is the population distribution and $p(\mathbf{\hat{y}}|\theta)$ is the likelihood of the data. Considering $N$ observations and the definitions of $\varphi$ and $\theta$,
\begin{equation}
    p(\mathbf{y}, \bm{\mu}, \bm{\Sigma} | \mathbf{\hat{y}}) \propto p(\bm{\mu}, \bm{\Sigma}) \prod_{j=1}^N p(\mathbf{y_j} | \bm{\mu}, \bm{\Sigma}) p(\mathbf{\hat{y}_j} | \mathbf{y_j})
    \label{eq:hier_base}
\end{equation}
where $p(\mathbf{\hat{y}_j} | \mathbf{y_j})\equiv p(\mathbf{\hat{y}_j} | \mathbf{y_j}, \mathbf{E_j})$ is the error distribution. Here, $\mathbf{E_j}$ is the observational error matrix for the $j$-th observation (see section \ref{sec:cov_errs}), which is responsible for covariant error propagation. Integrating out the true (latent) variable $\mathbf{y}$ gives,
\begin{align}
    p(\bm{\mu}, \bm{\Sigma} | \mathbf{\hat{y}}) &\propto \prod_{j=1}^N \int p(\mathbf{y_j} | \bm{\mu}, \bm{\Sigma}) p(\mathbf{\hat{y}_j} | \mathbf{y_j}, \mathbf{E_j}) d\mathbf{y_j}\nonumber \\
    &= p(\bm{\mu}, \bm{\Sigma}) \prod_{j=1}^N p(\mathbf{\hat{y}_j} | \bm{\mu}, \bm{\Sigma} + \mathbf{E_j})~,
    \label{eq:errconv}
\end{align}
where the integral should be understood as considering all possible values for the true (latent) variables. This process gives us the convolution of the error distribution with the model. Here, the Gaussian likelihood for a given galaxy $j$ can be given as $\mathcal{N}(\mathbf{\hat{y}_j} | \bm{\mu}, \bm{\Sigma}+\mathbf{E_j})$, which means,
\begin{align}
    p(\mathbf{\hat{y}_j}\,|\,&\bm{\mu},~ \bm{\Sigma}+\mathbf{E_j}) \nonumber\\
    &= \frac{ \exp{\left[ -\frac{1}{2} (\mathbf{\hat{y}_j}-\bm{\mu}) (\bm{\Sigma}+\mathbf{E_j})^{-1} (\mathbf{\hat{y}_j}-\bm{\mu})^\intercal \right]} }{(2\pi)^{M/2} \sqrt{|\bm{\Sigma}+\mathbf{E_j}|}}.
    \label{eq:mdgauss}
\end{align}
Equations~(\ref{eq:errconv}) and (\ref{eq:mdgauss}) constitute the Bayesian $M$D generalisation of 3D Gaussian ML method.

\subsection{Covariant error propagation}\label{sec:cov_errs}

Due to measurement errors, each galaxy in the observed $\mathbf{\hat{y}_j}$ has an uncertainty $\varepsilon_j$. As mentioned in the beginning of this section and just like the ML method used in \cite{magoulas2012, springob2014} and \cite{khaled2020}, Gaussian approach enables us to easily account for the errors along with their possible covariances. This is possible by incorporating uncertainties into the error matrix for each galaxy and these matrices will \textit{not} be diagonal due to covariant errors. 

For a data set such as $\mathbf{\hat{y}}=(\bm{r},\bm{s},\bm{i},\bm{m^\ast},\bm{c},\bm{\nu})$, the error matrix $\mathbf{E_j}$ can be constructed as,
\begin{equation}
    \mathbf{E_j}=\begin{pmatrix} 
                    \varepsilon_{r_j}^2 & 0 & \varepsilon_{r_j,i_j} & 0 & 0 & \varepsilon_{r_j, \nu_j} \\ 
                    0 & \varepsilon_{s_j}^2 & 0 & 0 & 0 & 0\\
                    \varepsilon_{r_j, i_j} & 0 & \varepsilon_{i_j}^2 & 0 & 0 & \varepsilon_{i_j, \nu_j}\\
                    0 & 0 & 0 & \varepsilon_{m^\ast_j}^2 & \varepsilon_{m^\ast_j, c_j} & 0\\
                    0 & 0 & 0 & \varepsilon_{m^\ast_j, c_j} & \varepsilon_{c_j}^2 & 0 \\
                    \varepsilon_{r_j, \nu_j} & 0 & \varepsilon_{i_j, \nu_j} & 0 & 0 & \varepsilon_{\nu_j}^2 \label{eq:fullerrormatrix}
    \end{pmatrix}
\end{equation}
where $\varepsilon_{x_j}$ is the uncertainty of the observable parameter $x$ for $j-$th galaxy. The off-diagonal elements represent the covariances between the errors of parameter pairs as $\varepsilon_{x_j, y_j}=\rho_{xy}^\varepsilon \varepsilon_{x_j}\varepsilon_{y_j}$ where $\rho_{xy}^\varepsilon$ is the correlation coefficient of the errors in $x$ and $y$ parameters. Calculation of $\rho_{xy}^\varepsilon$ is detailed in Appendix \ref{sec:calc_coverrors}.

\subsection{Selection effects}\label{sec:selection_effects}

Due to selection effects, the data $\mathbf{\hat{y}}$ is a truncated Gaussian, thus, the likelihood $p(\mathbf{\hat{y}_j} | \bm{\mu}, \bm{\Sigma} + \mathbf{E_j})$ needs to be renormalised by,
\begin{align}
    & p(\mathbf{\hat{y}_j} | \bm{\mu}, \bm{\Sigma} + \mathbf{E_j}) \rightarrow \frac{p(\mathbf{\hat{y}_j} | \bm{\mu}, \bm{\Sigma} + \mathbf{E_j})}{\hat{f}_j}, \nonumber\\
    & \hat{f}_j = \int\limits_{\mathbf{\hat{y}}_{\text{cut}}}^\infty p(\mathbf{\hat{y}_j} | \bm{\mu}, \bm{\Sigma} + \mathbf{E_j}) d^M\mathbf{\hat{y}_j}
    \label{eq:fj_conv}
\end{align}
where $\hat{f}_j$ is the normalisation integral for the multivariate normal distribution and $\bm{\hat{y}_\text{cut}}$ is the vector corresponding to the lower limits of the parameters included in the $\mathbf{\hat{y}}$ vector. $\hat{f}_j$ ensures that the likelihood $\int p(\mathbf{\hat{y}_j} | \bm{\mu}, \bm{\Sigma} + \mathbf{E_j}) d^M \mathbf{\hat{y}_j} = 1$ within the domain defined by the data. Due to the instrumental limits, we expect to see explicit cuts in velocity dispersion (or $M_\star$, as discussed in section \ref{sec:data_sample}), apparent magnitude and redshift. The redshift and magnitude limits can be simultaneously corrected by \textit{selection function}, $S$, defined in \cite{magoulas2012} which is inversely proportional to $V_\text{max}$ weighting \citep[][]{schmidt1968}: $S\propto1/V_\text{max}$. When the weight $w_j=1/S_j$ is applied to the renormalised likelihood as $[ p(\mathbf{\hat{y}_j} | \bm{\mu}, \bm{\Sigma} + \mathbf{E_j})/\hat{f}_j]^{w_j}$, the natural logarithm of equation~(\ref{eq:errconv}) becomes,

\begin{align}
    \ln p(\bm{\mu}, \bm{\Sigma}\,&|\,\mathbf{\hat{y}}) \propto \ln p(\bm{\mu}, \bm{\Sigma}) \nonumber\\ 
    & + \sum_{j=1}^N w_j \left[ \ln p(\mathbf{\hat{y}_j} | \bm{\mu}, \bm{\Sigma} + \mathbf{E_j}) - \ln \hat{f}_j \right].
    \label{eq:log_conv}
\end{align}

An important aspect of weighting here is that the $w_j$ weights are small for our sample: $w_j=1$ for 2793 galaxies (i.e., no weighting needed at all) out of 2850 and maximum weight encountered is $w_j=2.54$. This maximum value of the weights is minuscule especially considering that \cite{taylor2011} and \cite{magoulas2012} have chosen their samples discarding the galaxies with too large weights, $w>30$ and $w>20$ respectively, to prevent these galaxies from skewing the fits. The reason we obtain such low weights is the fact that GAMA provides a deep probe of the massive galaxy population with $\log M_\star\gtrsim 10.5$ out to $z\approx 0.25$, as pointed out in \cite{taylor2011}, which is the bulk of our sample. 

The log-likelihood in equation~(\ref{eq:log_conv}) enables sampling from a truncated and censored multivariate Gaussian, thus, accounts for the selection effects and provides an unbiased fit.

\subsection{Outlier rejection: mixture model}\label{sec:mixture}

Conventional weighting of data points with the inverse square of their observational uncertainties $(w_j=1/\varepsilon_j^2)$ is prone to be sensitive to outliers \citep[e.g.][]{hogg2010, taylor2015}. A proposed solution to this \textit{`bad data'} problem comes from Gaussian mixture modelling, in which a secondary component is added to the likelihood function. This enables us to model the outliers as some fraction of the data being drawn from a \textit{bad distribution}, whereas the \textit{`good data'} that we are interested in is drawn from the \textit{good distribution}.

We model this secondary component as a Gaussian, because we simply do not have any knowledge of the bad distribution. A key point here is that no a priori assumptions are made as to the goodness and/or badness of the data points so that the model can treat every point as having an equal probability of being bad, thus, it can objectively identify the outliers \citep{taylor2015}. If the fraction of good data is denoted as $f_\text{good}$, the net likelihood becomes,
\begin{align}
    p_\text{net}(\mathbf{\hat{y}} | \bm{\mu}, \bm{\Sigma}, f_\text{good}) &= \prod_{j=1}^N [ f_\text{good} p(\mathbf{\hat{y}_j}|\bm{\mu}, \bm{\Sigma}) \nonumber \\
    &+ (1-f_\text{good})p(\mathbf{\hat{y}_j}|\bm{\mu}, \bm{\Sigma}_\text{bad}) ]
    \label{eq:netlikelihood}
\end{align}
where $\bm{\Sigma}_\text{bad}$ is the covariance matrix of the bad distribution. It should also be noted that both good and bad distributions here are assumed to have the same mean $\bm{\mu}$. Equation~(\ref{eq:netlikelihood}) shows that each data point $\mathbf{\hat{y}_j}$ may have arisen from either good or bad distribution. Inserting equation~(\ref{eq:netlikelihood}) to the posterior in equation~(\ref{eq:log_conv}) we obtain the log-likelihood:
\begin{align}
    \ln p(\bm{\mu}, \bm{\Sigma}, f_\text{good} | \mathbf{\hat{y}}) & \propto \sum_{j=1}^N \ln \Biggl[  f_\text{good}\left(\frac{p(\mathbf{\hat{y}_j}|\bm{\mu}, \bm{\Sigma}+\mathbf{E_j})}{\hat{f}_j}\right)^{w_j} \nonumber\\
    &+ (1-f_\text{good})p(\mathbf{\hat{y}_j}|\bm{\mu}, \bm{\Sigma}_\text{bad})\Biggr].
    \label{eq:wbc_mix}
\end{align}
It is noteworthy that, as seen from equation~(\ref{eq:wbc_mix}), $f_\text{good}$ is also declared as a parameter. Furthermore, we can quantify the badness of each data point, i.e., the probability of a data point being bad, by $f_\text{bad}p_\text{bad, j}/p_\text{net, j}$ where $f_\text{bad}=1-f_\text{good}$. One final remark is that we also assume the bad distribution is not truncated.

\subsection{STAN implementation}\label{sec:stan}

Within this study, we typically use MCMC samples from 4 Markov chains with each chain consisting of 1000 draws from the posterior in equation~(\ref{eq:wbc_mix}), 500 of which is warm up, thus adding up to 2000 iterations in total, after discarding the warm up. Within this Bayesian MCMC framework, it is crucial to implement convoluted statistical models, like the one we present in this work, in an efficient way to avoid computationally expensive practices. To do this, we make use of optimised and robust built-in functions of \textit{STAN}.

The critical and possibly the most computationally challenging part of applying equation~(\ref{eq:wbc_mix}) is the normalisation factor $\hat{f}_j$ which accounts for the selection cuts in velocity dispersion and stellar mass. We give the derivation of an expression for $\hat{f}_j$ in Appendix \ref{sec:fjcalc} in terms of the built-in cumulative and probability distribution functions.

Another challenge arises from the Gaussian mixture. As seen from equation~(\ref{eq:wbc_mix}), the log-posterior is the natural logarithm of the sum of two terms which can be handled by one of the mainstays of statistical computing, log-sum of exponentials, where $\ln(a+b)=\ln[\exp{(\ln a)} + \exp{(\ln b)}] = \text{log-sum-exp}(\ln a, \ln b)$. This way, equation~(\ref{eq:wbc_mix}) becomes
\begin{align}
    \ln & p(\bm{\mu}, \bm{\Sigma}, f_\text{good} | \mathbf{\hat{y}}) \propto \nonumber \\
    & \sum_{j=1}^N \text{log-sum-exp}\biggl( \ln f_\text{good} + w_j [\ln p(\mathbf{\hat{y}_j}|\bm{\mu}, \bm{\Sigma}+\mathbf{E_j}) \nonumber \\ 
    &- \ln \hat{f}_j], \ln(1-f_\text{good}) + \ln p(\mathbf{\hat{y}_j}|\bm{\mu}, \bm{\Sigma}_\text{bad}) \biggr)
    \label{eq:wbc_logsumexp}
\end{align}
which makes efficient sampling possible by using \textit{STAN}'s optimised functions \verb|log_sum_exp| and \verb|log1m|$(x)=\ln(1-x)$. Equation~(\ref{eq:wbc_logsumexp}) is the final form for the posterior of our model from which the MCMC samples will be drawn in \textit{STAN}.

\subsection{Extracting relations between galaxy properties: linear transformations}\label{sec:lintransform}

All information regarding the relations between the main galaxy parameters in $\mathbf{\hat{y}}$, and other parameters that can be derived from them, are encoded in the covariance matrix, $\bm{\Sigma}$. Relations between parameters of interest can be extracted from $\bm{\Sigma}$ using linear transformations,
\begin{align}
    T: \mathbb{R}^M &\longrightarrow \mathbb{R}^K\nonumber\\
    \bm{y} &\longrightarrow \bm{\omega}=\bm{Ay}+\bm{B}
    \label{eq:lintransform}
\end{align}
where $K$ is the number of parameters of interest. The Jacobian of this transformation $,\bm{J}=\bm{A}\in\mathbb{R}^{K\times M}$, can then be used in,
\begin{equation}
    \bm{\Sigma}_\omega = \bm{A}\bm{\Sigma}\bm{A}^\intercal, \qquad \bm{\bar{\omega}}=\bm{A\bar{y}}+\bm{B}
    \label{eq:newcovmean}
\end{equation}
to find the covariance matrix $(\bm{\Sigma}_\omega \in \mathbb{R}^{K\times K})$ and the mean vector $(\bm{\bar{\omega}} \in \mathbb{R}^K)$ for those parameters. The ordinary least squares (OLS) coefficients for a linear relation between these $K$ parameters can then be calculated from $\bm{\Sigma}_\omega$ and $\bm{\bar{\omega}}$. Furthermore, orthogonal distances regression (ODR) coefficients can also be obtained from the eigenvectors of $\bm{\Sigma}_\omega$.

\subsection{Likelihood Normalisation}\label{sec:fjcalc}

In this section, we show the main steps to derive an expression for the integral normalisation factor $\hat{f}_j$, similar to the works of \cite{magoulas2012} and \cite{dam2020}. For the data set $\mathbf{\hat{y}}=(\bm{r}, \bm{s}, \bm{i}, \bm{m^*}, \bm{\ell}, \bm{c}, \bm{\nu}, \bm{m^d})$, the selection limits in our final sample are $m^*\geqslant m_\text{cut}=10.3$, $60<\sigma_e\text{[km/s]}<450$, $0.01\leqslant z \leqslant 0.12$ and $m_r\leqslant19.65$, as discussed in section \ref{sec:data}. Since magnitude and redshift limits can be accounted for by selection probability function $S\propto1/V_\text{max}$, we can say that there are no cut-off values for any component of $\mathbf{\hat{y}}$ except for $\bm{m^*}$ and $\bm{s}$ which means that the integral over all the other components will just be the marginalisation of the 8D probability distribution function (PDF). In this case, the likelihood normalisation factor $f_j$ in equation~(\ref{eq:wbc_logsumexp}) can be expressed in terms of a bivariate normal PDF,
\begin{align}
    f_j &= \int\limits_{m_{\text{cut}}}^\infty \int\limits_{s_\text{low}}^{s_\text{up}} p(\bm{\hat{y}^\prime_j} | \bm{\mu^\prime}, \bm{C_j^\prime}) d^2\bm{\hat{y}^\prime_j} \nonumber \\ 
    &= \int\limits_{m_{\text{cut}}}^\infty \int\limits_{s_\text{low}}^{s_\text{up}} \frac{ \exp{ \left[ -\frac{1}{2} (\bm{\hat{y}^\prime_j} - \bm{\mu^\prime})^\intercal (\bm{C_j^\prime})^{-1} (\bm{\hat{y}^\prime_j} - \bm{\mu^\prime})\right]} }{2\pi \sqrt{|\bm{C_j^\prime}}|} ds_j dm_j
    \label{eq:fj_bvn}
\end{align}
where $\bm{\hat{y}^\prime_j} = (s_j, m^*_j)$, mean vector $\bm{\mu^\prime}=(\bar{s}, \bar{m}^*)$ and $\bm{C_j^\prime}$ is constructed with the $s, m$ elements of the covariance matrix convolved with the error matrix, which can be done via
\begin{align}
    \bm{C_j^\prime} = \bm{J^\prime} (\bm{\Sigma}+\bm{E_j}) \bm{{J^\prime}}^\intercal,\quad \bm{J^\prime} = \begin{pmatrix}
        0 & 1 & 0 & 0 & 0 & 0 & 0 & 0 \\
        0 & 0 & 0 & 1 & 0 & 0 & 0 & 0 
    \end{pmatrix}.
\end{align}
Using the Cholesky decomposition of $\bm{C_j^\prime} = L_j L_j^\intercal$, where $L_j$ is a lower triangular matrix, provides an easier way to reduce equation~(\ref{eq:fj_bvn}) to a more applicable form for efficient MCMC sampling in \textit{STAN}. This way, equation~(\ref{eq:fj_bvn}) can be expressed as,
\begin{align}
    &f_j = \int\limits_{u_{j,\text{cut}}}^\infty \phi(u_j)\left[ \Phi(\alpha_\text{up} + \beta u_j) - \Phi(\alpha_\text{low} + \beta u_j) \right] du_j \nonumber \\
    &\text{where, } u_j = \frac{m^*_j-\bar{m}^*}{L_{j,11}},\quad u_{j,\text{cut}} = \frac{m_\text{cut}-\bar{m}^*}{L_{j,11}}, \nonumber \\
    & \alpha_\text{low} = \frac{s_\text{low} - \bar{s}}{L_{j,22}}, \quad \alpha_\text{up} = \frac{s_\text{up} - \bar{s}}{L_{j,22}} \text{ and } \beta = - \frac{L_{j,21}}{L_{j,22}}
\end{align}
Here, $\Phi$ is the cumulative distribution function (CDF) and $\phi$ is the PDF of the standard normal distribution, having the relation,
\begin{equation}
    \Phi(x) = \frac{1}{\sqrt{2\pi}}\int\limits_{-\infty}^x e^{-t^2/2} dt = \int\limits_{-\infty}^x \phi(t) dt \text{ and } \phi(x) = \frac{d\Phi(x)}{dx}.
\end{equation}
Using the tables provided by \cite{owen1980}, 
\begin{align}
    f_j &= \Phi\left( \frac{\alpha}{\sqrt{1+\beta^2}} \right)\nonumber \\ 
    &- \text{BvN}\left( \frac{\alpha}{\sqrt{1+\beta^2}}, u_{j,\text{cut}}, \frac{-\beta}{\sqrt{1+\beta^2}} \right) \Bigg|_{\alpha=\alpha_\text{low}}^{\alpha=\alpha_\text{up}}
    \label{eq:fj_smcut_final}
\end{align}
where BvN is the CDF of the standard bivariate normal distribution given as
\begin{align}
    \text{BvN}(h, k, \rho) &= \frac{1}{2\pi\sqrt{1 - \rho^2}} \nonumber \\ 
    & \times \int\limits_{-\infty}^k \int\limits_{-\infty}^h \text{exp}\left[ -\frac{1}{2}\left( \frac{x^2 - 2\rho xy + y^2}{1 - \rho^2}\right) \right]dx dy.
\end{align}
We use the algorithm of \cite{richardj1989} to evaluate BvN, also making use of \textit{STAN}'s optimised built-in functions for $\Phi$ and $\phi$.

\section{The stellar-to-dynamical mass relation with different structure correction factors}\label{sec:homology}

For assessing the effects of the homology assumption, we additionally inspect the relation between $M_\star$ and $\sigma_e^2 R_e$, i.e., the dynamical mass calculated with $k=$ constant. To do this, we extract the relation between $M_\star, \sigma_e, R_e$ by constructing the conditional $(\bm{m^*, 2s+r | s, \nu, c})$ and marginal $(\bm{m^*, 2s+r})$ distributions from our model of the 8D parameter space. We use the linear transformation $\bm{Ay+B}$ where
\begin{equation}
    \bm{A}_\text{cond} = 
    \begin{pmatrix}
        0 & 0 & 0 & 1 & 0 & 0 & 0 & 0 \\
        1 & 2 & 0 & 0 & 0 & 0 & 0 & 0 \\
        0 & 1 & 0 & 0 & 0 & 0 & 0 & 0 \\
        0 & 0 & 0 & 0 & 0 & 1 & 0 & 0 \\
        0 & 0 & 0 & 0 & 0 & 0 & 1 & 0 
    \end{pmatrix}
\end{equation}
for the conditional distribution and $\bm{B}$ encapsulates the constants arising from $k/G$ and unit conversions. For the marginal distribution, we just take the first two rows of $\bm{A}_\text{cond}$. Applying these transformations yield the results given in Table \ref{tab:cond_marg_k5}.
\begin{table*}
  \centering
  \caption{Same as Table \ref{tab:cond_marg}, but with the assumption of homology, i.e., $G M_\text{dyn}=k \sigma_e^2 R_e$ with $k=5$.}
       \resizebox{0.7\textwidth}{!}{
    \begin{tabular}{crccrcc}
        \cmidrule{3-7} &   & \multicolumn{2}{c}{$M_\star\propto (\sigma_e^2 R_e)^\alpha$} &   & \multicolumn{2}{c}{$(\sigma_e^2 R_e)\propto M_\star^\beta$} \\
        \cmidrule{3-7} &   & Quiescent & Star-forming &   & Quiescent & Star-forming \\
            \midrule
        \multirow{5}[2]{*}{Conditional} & $\alpha^\perp$ & $0.976\pm0.008$ & $1.278\pm0.024$ & $\beta^\perp$ & $1.025\pm0.009$ & $0.782\pm0.015$ \\
          & $\alpha^\parallel$ & $0.846\pm0.007$ & $0.809\pm0.014$ & $\beta^\parallel$ & $0.882\pm0.007$ & $0.577\pm0.01$ \\
          & $\rho$ & $0.864\pm0.003$ & $0.683\pm0.008$ & - & - & - \\
          & $\sigma_\alpha^\perp$ & $0.057\pm0.001$ & $0.063\pm0.001$ & $\sigma_\beta^\perp$ & $0.056\pm0.001$ & $0.081\pm0.001$ \\
          & $\sigma_\alpha^\parallel$ & $0.108\pm0.001$ & $0.147\pm0.002$ & $\sigma_\beta^\parallel$ & $0.11\pm0.001$ & $0.124\pm0.001$ \\
        \midrule
        \multirow{5}[2]{*}{Marginal} & $\alpha^\perp$ & $0.773\pm0.004$ & $0.822\pm0.01$ & $\beta^\perp$ & $1.294\pm0.007$ & $1.217\pm0.014$ \\
          & $\alpha^\parallel$ & $0.743\pm0.004$ & $0.748\pm0.009$ & $\beta^\parallel$ & $1.214\pm0.007$ & $1.062\pm0.012$ \\
          & $\rho$ & $0.95\pm0.001$ & $0.891\pm0.004$ & - & - & - \\
          & $\sigma_\alpha^\perp$ & $0.073\pm0.001$ & $0.092\pm0.001$ & $\sigma_\beta^\perp$ & $0.056\pm0.001$ & $0.076\pm0.001$ \\
          & $\sigma_\alpha^\parallel$ & $0.115\pm0.001$ & $0.152\pm0.002$ & $\sigma_\beta^\parallel$ & $0.147\pm0.001$ & $0.181\pm0.002$ \\
        \bottomrule
        \bottomrule
    \end{tabular}
    }
  \label{tab:cond_marg_k5}
\end{table*}

After testing the homology assumption by taking $k=5$, we proceed to analyse the case of $k(n)$ of \cite{cappellari2006}. We give the corresponding results in Table \ref{tab:kn_cappellari} in the same way as Tables \ref{tab:cond_marg} and \ref{tab:cond_marg_k5}.
\begin{table*}[htbp]
  \centering
  \caption{Same as Tables \ref{tab:cond_marg} and \ref{tab:cond_marg_k5} but for when $M_\mathrm{dyn}$ is calculated with the structure correction factor of \protect\cite{cappellari2006} in equation (\ref{eq:bn}).}
       \resizebox{0.7\textwidth}{!}{
    \begin{tabular}{crccrcc}
\cmidrule{3-7}      &   & \multicolumn{2}{c}{$M_\star\propto M_\text{dyn}^\alpha$} &   & \multicolumn{2}{c}{$M_\text{dyn}\propto M_\star^\beta$} \\
\cmidrule{3-7}      &   & Quiescent & Star-forming &   & Quiescent & Star-forming \\
    \midrule
    \multirow{5}[2]{*}{Conditional} & $\alpha^\perp$ & $1.013\pm0.03$ & $1.4\pm0.093$ & $\beta^\perp$ & $0.988\pm0.029$ & $0.718\pm0.047$ \\
      & $\alpha^\parallel$ & $0.87\pm0.025$ & $0.822\pm0.053$ & $\beta^\parallel$ & $0.852\pm0.025$ & $0.527\pm0.034$ \\
      & $\rho$ & $0.861\pm0.011$ & $0.658\pm0.031$ & - & - & - \\
      & $\sigma_\alpha^\perp$ & $0.06\pm0.002$ & $0.063\pm0.003$ & $\sigma_\beta^\perp$ & $0.061\pm0.002$ & $0.087\pm0.004$ \\
      & $\sigma_\alpha^\parallel$ & $0.117\pm0.004$ & $0.158\pm0.007$ & $\sigma_\beta^\parallel$ & $0.116\pm0.004$ & $0.126\pm0.004$ \\
    \midrule
    \multirow{5}[2]{*}{Marginal} & $\alpha^\perp$ & $0.828\pm0.016$ & $0.86\pm0.036$ & $\beta^\perp$ & $1.208\pm0.023$ & $1.164\pm0.048$ \\
      & $\alpha^\parallel$ & $0.788\pm0.014$ & $0.724\pm0.03$ & $\beta^\parallel$ & $1.125\pm0.022$ & $0.929\pm0.037$ \\
      & $\rho$ & $0.942\pm0.005$ & $0.82\pm0.018$ & - & - & - \\
      & $\sigma_\alpha^\perp$ & $0.077\pm0.002$ & $0.106\pm0.004$ & $\sigma_\beta^\perp$ & $0.064\pm0.002$ & $0.092\pm0.005$ \\
      & $\sigma_\alpha^\parallel$ & $0.129\pm0.005$ & $0.179\pm0.008$ & $\sigma_\beta^\parallel$ & $0.154\pm0.005$ & $0.202\pm0.006$ \\
    \bottomrule
    \bottomrule
    \end{tabular}
    }
  \label{tab:kn_cappellari}
\end{table*}

\begin{figure*}
    \centering
    \includegraphics[width=\textwidth]{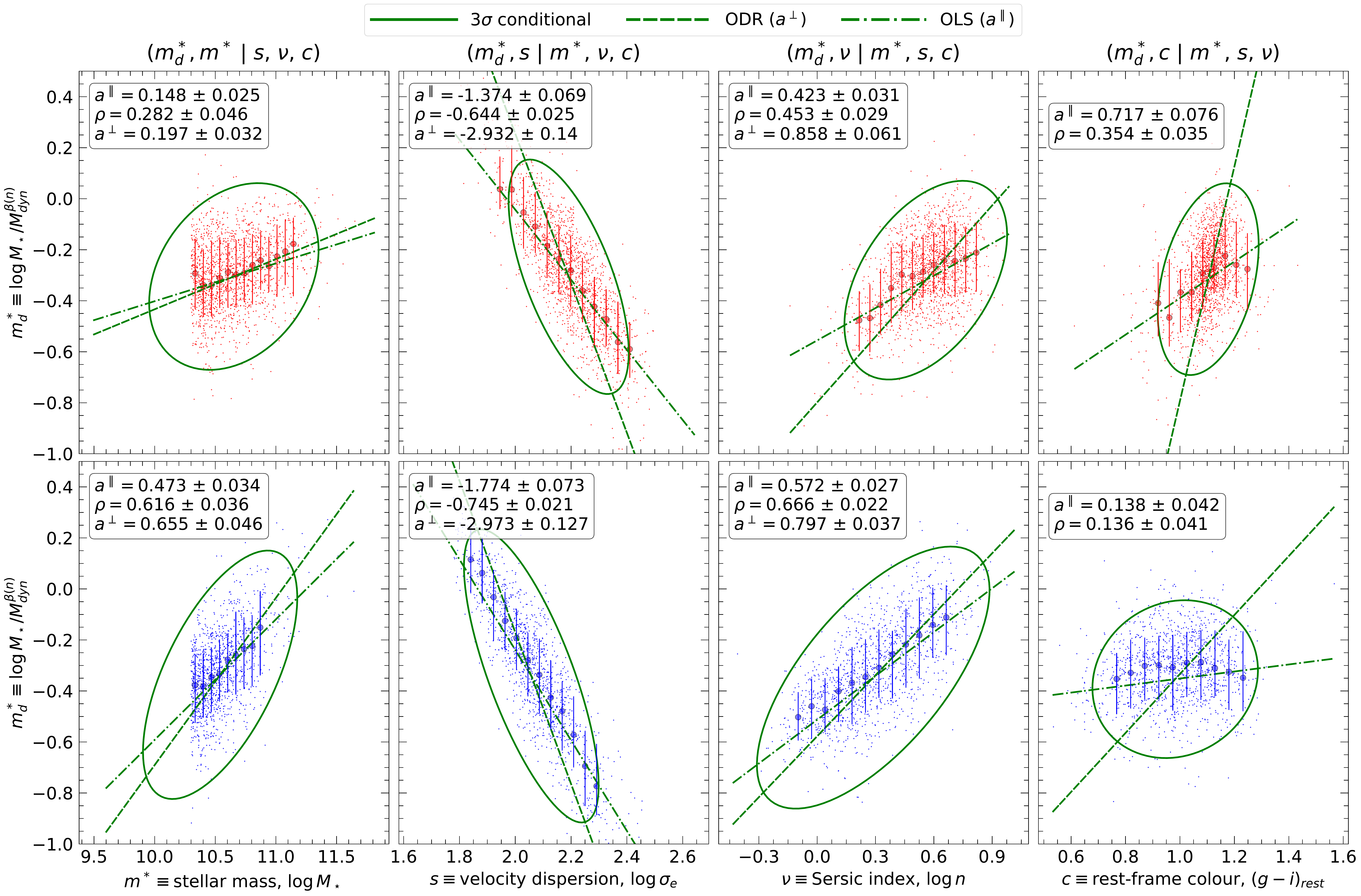}
    \caption{Same as Figure \ref{fig:msmd_conditional}, but for when $M_\mathrm{dyn}$ is calculated with the structure correction factor of \protect\cite{cappellari2006} in equation (\ref{eq:bn}). We denote this with $M_\mathrm{dyn}^{\beta(n)}$ because the $k(n)$ factor is denoted with $\beta(n)$ in \protect\cite{cappellari2006}.}
    \label{fig:msmd_trends_cappellari}
\end{figure*}

Finally, in Figure \ref{fig:msmd_trends_cappellari}, we show the dependence of $M_\star/M_\mathrm{dyn}$ relation on $m^*, s, \nu$ and $(g-i)_\mathrm{rest}$ when $k(n)$ from \cite{cappellari2006} is used.

\section{Additional discussion on SPS-related potential systematics}\label{sec:more_discussion}

In this appendix, we present the more detailed results of fitting the $\log\,M_\star/\hat{M}_\star$ residuals as a function of SP parameters shown in Figure \ref{fig:mstar_sps_residuals}, listed in Table \ref{tab:spsparams}. We also discuss a few tentative implications of these results.

\begin{table*}
  \centering
  \caption{Summary of the linear fits in the form $\Delta \log M_\star=Ax+B$ inferred from the 2D-Gaussian models of the relations between the residuals and stellar population parameters, for both Q and SF galaxies. Errors are derived from the posteriors. The implied scatter given under the column $\sigma_{x, \text{implied} }=A\sigma_x$ for each $x-$quantity is the amount of scatter in $\log M_\star$ which can be linked to that quantity. The value that tells us what fraction of intrinsic scatter corresponds to the quantity in question is given under the column $f_{x, \text{int}}$.}
   \resizebox{\textwidth}{!}{
      \begin{tabular}{lcccccccccc}
    \cmidrule{2-11}    & \multicolumn{5}{c}{Quiescent} & \multicolumn{5}{c}{Star Forming} \\
      \midrule
      Quantity, $x$ & A & $\sigma_x$ & $\rho$ & $\sigma_\text{implied}$ & $f_{x, \text{int}}$ & A & $\sigma_x$ & $\rho$ & $\sigma_\text{implied}$ & $f_{x, \text{int}}$ \\
      \midrule
      $(g-i)$ & $0.000\pm0.026$ & $0.108\pm0.001$ & $0.000\pm0.014$ & $0.000\pm0.003$ & 0.000 & $0.000\pm0.016$ & $0.147\pm0.002$ & $0.000\pm0.016$ & $0.000\pm0.002$ & 0.000 \\
      $\sinh^{-1}(H\alpha_\text{EW})$ & $-0.013\pm0.006$ & $0.557\pm0.011$ & $-0.058\pm0.029$ & $0.007\pm0.004$ & 0.005 & $0.011\pm0.005$ & $0.922\pm0.019$ & $0.068\pm0.031$ & $0.010\pm0.005$ & 0.005 \\
      $H\delta_\text{EW}$ & $0.021\pm0.003$ & $1.238\pm0.029$ & $0.207\pm0.031$ & $0.026\pm0.004$ & 0.058 & $0.011\pm0.003$ & $1.682\pm0.043$ & $0.130\pm0.035$ & $0.019\pm0.005$ & 0.017 \\
      $D_n 4000$ & $-0.108\pm0.026$ & $0.141\pm0.003$ & $-0.122\pm0.029$ & $0.015\pm0.004$ & 0.020 & $-0.113\pm0.028$ & $0.169\pm0.004$ & $-0.130\pm0.032$ & $0.019\pm0.005$ & 0.017 \\
      $E(B-V)$ & $-1.178\pm0.107$ & $0.058\pm0.003$ & $-0.538\pm0.040$ & $0.068\pm0.006$ & 0.394 & $-0.323\pm0.134$ & $0.060\pm0.003$ & $-0.130\pm0.053$ & $0.019\pm0.008$ & 0.017 \\
      $\log\,$ sSFR (Gyr$^{-1}$) & $-0.107\pm0.033$ & $0.147\pm0.005$ & $-0.126\pm0.038$ & $0.016\pm0.005$ & 0.021 & $-0.023\pm0.016$ & $0.312\pm0.008$ & $-0.048\pm0.034$ & $0.007\pm0.005$ & 0.002 \\
      $\log\langle t_\star\rangle_\text{LW}$ (Gyr) & $0.205\pm0.023$ & $0.204\pm0.006$ & $0.322\pm0.034$ & $0.042\pm0.005$ & 0.149 & $0.306\pm0.081$ & $0.113\pm0.007$ & $0.222\pm0.057$ & $0.035\pm0.009$ & 0.055 \\
      $\log\,Z_\star/Z_\odot$ & $0.029\pm0.037$ & $0.207\pm0.011$ & $0.048\pm0.061$ & $0.006\pm0.008$ & 0.003 & $-0.080\pm0.037$ & $0.264\pm0.014$ & $-0.144\pm0.066$ & $0.021\pm0.010$ & 0.021 \\
      \bottomrule
      \end{tabular}%
    }
  \label{tab:spsparams}
\end{table*}

It is noteworthy that $\sim 2\%$ of the mismatch seems to be stemming from sSFR for quiescent galaxies, while it has no contribution at all for the star-forming galaxies, presenting a somewhat interesting aspect. In an attempt to explain this, we should simultaneously examine the distributions of stellar population parameters and their relations with each other. In Figure \ref{fig:sps_gama}, for our GAMA sample, we provide these distributions for the SP parameters listed in Table \ref{tab:spsparams} and additionally $\log M_\star$. 

The $D_n(4000)-H\delta$ plane can be used as an observational diagnostic tool for the recent star formation histories (SFHs) of galaxies where galaxies which have undergone recent starburst events that ended $\sim 0.1-1$ Gyr ago will exhibit the strongest $H\delta$ absorption lines \citep[e.g.][]{kauffmann2003a, gallazzi2005, gallazzi2009}. As seen from the $H\delta-\log M_\star$, $D_n(4000)-H\delta$ and $H\delta-\log\langle t_\star\rangle_\text{LW}$ panels of Figure \ref{fig:sps_gama}, the strongest $H\delta$ absorption lines are seen in quiescent galaxies with the highest mass, strongest 4000\AA break and oldest stellar populations, thus, suggesting that some of the quiescent galaxies in our sample, most likely the ones with the highest masses, might have gone through recent starburst episodes. However, as detailed in \cite{taylor2011}, stellar masses in GAMA were obtained with single component star formation history [SFH; $\psi_\star(t)$], i.e., assuming a smooth and exponentially declining SFH which means that the SFH does not include starburst component(s). The usage of such single component SFH models introduces a fundamental limit to SED modelling in which young stars formed during a recent episode of starburst will be so bright that they outshine the older stars with lower masses which in turn causes the $M_\star$ and SFRs to be underestimated, while the luminosity weighted mean stellar ages are overestimated \citep[e.g.][]{papovich2001, lee2009b}. This might be the cause of the opposite trends in sSFR and stellar age for quiescent galaxies.

According to Figure \ref{fig:mstar_sps_residuals}, the underestimated $M_\star$ and sSFRs of the Qs lie in the interval $-1.5 \lesssim \log s\text{SFR}\lesssim -1$, corresponding to intrinsic colours in the interval $0.6\lesssim (g-i)_\star \lesssim 0.8$ (Figure \ref{fig:sps_gama}) and leading to the overestimation of $(g-i)_\star$ in that interval. These lead to the overestimation of $D_n 4000 \gtrsim 1.8$ and underestimation of H$_\delta \lesssim -1$, potentially providing a qualitative explanation to the residual trends seen with these spectral indices. These results align well with \cite{conroy2010b} who concluded that if Qs are assumed to be comprised of solely old and metal-rich stellar populations, all SPS models yield redder $ugr$ colours, stronger $D_n 4000$ and weaker $H_\delta$, hence the opposite trends seen between these properties in Figure \ref{fig:mstar_sps_residuals}.

\section{Stellar population parameters in GAMA}
\begin{figure*}
    \centering
    \includegraphics[width=\textwidth]{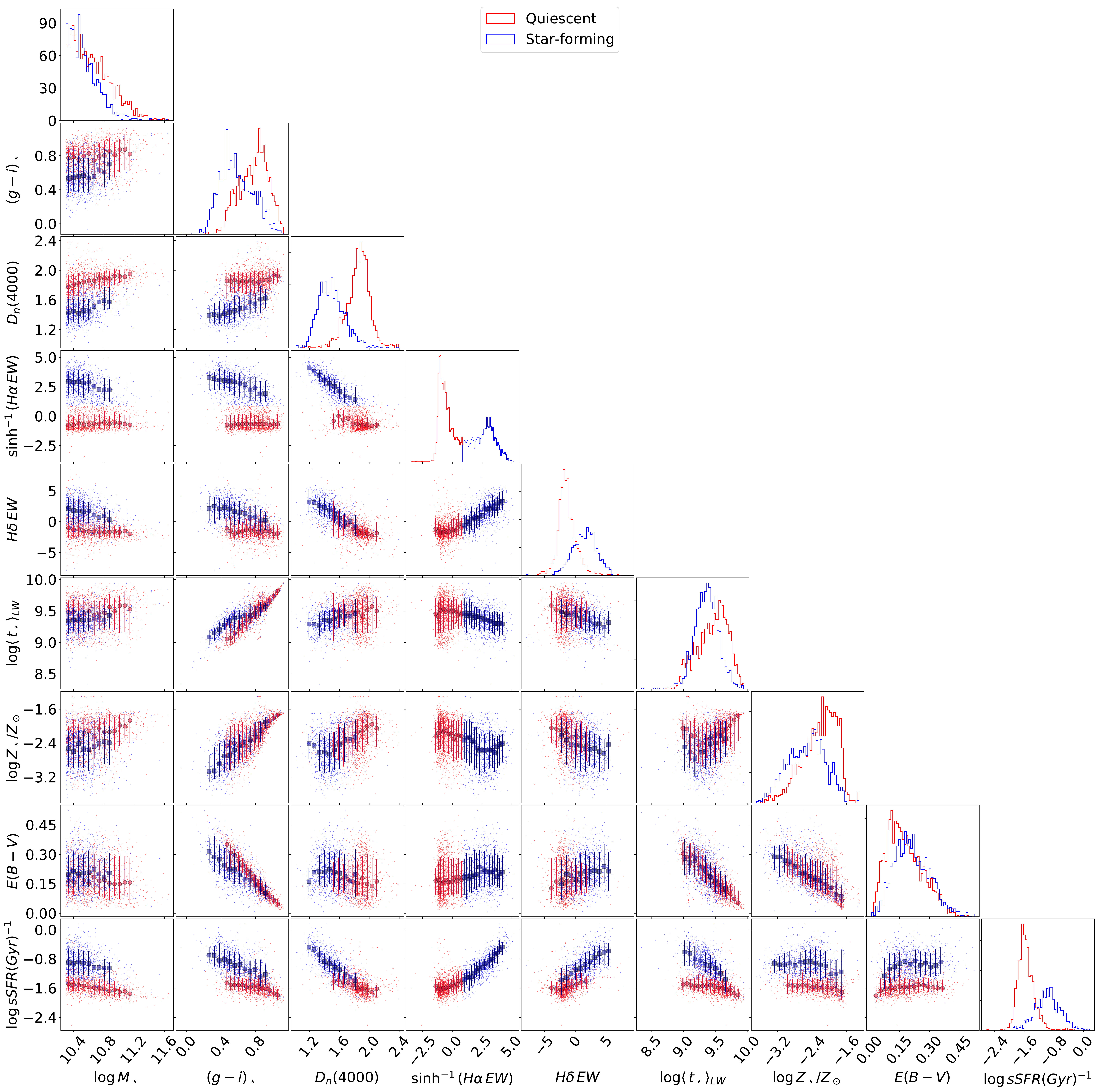}
    \caption{Distribution of stellar population parameters in the GAMA sample used in this study. Symbols and colours are the same as Figure \ref{fig:mstar_hyp}}
    \label{fig:sps_gama}
\end{figure*}

Here, in Figure \ref{fig:sps_gama}, we present the distributions of the stellar population parameters across our sample. Note that except for the spectral indices H$\alpha$, H$\delta$ and $D_n(4000)$, the parameters have been derived through SPS-modelling following \cite{taylor2011}.

\bibliography{refs}{}
\bibliographystyle{aasjournal}



\end{document}